\documentclass[a4paper,12pt]{article}
\usepackage[T1]{fontenc}
\usepackage[letterpaper,tmargin=2.5 cm,bmargin=2.5cm,rmargin=2cm,lmargin=2cm]{geometry}
\usepackage{amssymb,amsmath}
\usepackage{graphicx}
\usepackage{fancyhdr}
\usepackage{fancybox}
\usepackage{shadow}
\usepackage{empheq}
\usepackage{titlesec}
\usepackage{titletoc}
\usepackage{soul, color}
\usepackage{float}
\usepackage{shorttoc}
\usepackage{xcolor}
\usepackage{fancybox}
\usepackage{natbib}
\usepackage{array}
\usepackage{tabularx}
\usepackage{booktabs}
\usepackage{rotating}
\usepackage{bigstrut}
\usepackage{natbib}
\usepackage{aeguill}
\everymath{\displaystyle}
\usepackage{subfigure}

\newtheorem{prop}{Proposition}

\usepackage{changepage}

\title{Weak Identification and Estimation of Social Interaction Models\thanks{ \tiny{Comments from Marine Carrasco, Eric Renault, James G. MacKinnon, Silvia Goncalves, Russell Davidson, Yann Bramoull\'{e}, Lynda Khalaf, John Peirson and A. Colin Cameron are gratefully acknowledged. The author also
thanks seminar participants at the University of Kent, University of Ottawa, Universidad Pontificia Javeriana, RES 2016, EEA-ESEM 2016,
 AFES 2016, CIREQ Econometrics Conference 2017, the University of Bristol's Econometrics Study Group 2017, SCSE 2017 and NASM 2018 for their comments. Many thanks to  Xiaodong Lui for kindly providing his code for bias-corrected 2SLS methods and to  Teemu Lyytik{\"a}inen for providing data on Finnish municipalities.}}}
\author{Guy Tchuente\thanks{\tiny{E-mail: guytchuente@gmail.com. Address: School of Economics, University of Kent, Keynes College, Canterbury, Kent, CT27NP. Tel:+441227827249}}  
}

\date{ February 2019}
\begin{document}
\maketitle
\abstract{The identification of the network effect is based on either group size variation, the structure of the network or the relative position in the network. I provide easy-to-verify necessary conditions for identification of undirected network models based on the number of distinct eigenvalues of the adjacency matrix. Identification of network effects is possible; although in many empirical situations existing identification strategies may require the use of many instruments or instruments that could be strongly correlated with each other. The use of highly correlated instruments or many instruments may lead to weak identification or many instruments bias. This paper proposes regularized versions of the two-stage least squares (2SLS) estimators as a solution to these problems. The proposed estimators are consistent and asymptotically
normal. A Monte Carlo study illustrates the properties of the regularized estimators. An empirical application, assessing a local government tax competition model, shows the empirical relevance of using regularization methods.
 }
\newpage

\section{Introduction}
This paper investigates the estimation of social interaction  models with network structures
and the presence of endogenous, contextual, correlated and group fixed effects.\footnote{In network models, an agent's behavior may be influenced by his peers' choices (the endogenous effect), his peers' exogenous characteristics (the contextual effect), and/or by the common environment of the
network (the correlated effect) (see \cite{manski1993identification} for a description of these models).}
In his seminal paper on network model estimation, \cite{manski1993identification} argues that solving the reflection problem in identifying and estimating the endogenous interaction effects is of significant interest in social interaction models. He shows that the separate identification of the network effects, in a linear in mean model, is impossible. Following  \cite{manski1993identification},
the literature on identification of network effect has proposed three main identification strategies. They are based on either the variation in the size of the group of peers or on the structure through which peers interact. The present paper investigates a robust to weak identification estimation strategy. Weak identification can occur in limit cases for all the identification strategies.

The first method for identification was proposed by \cite{lee2007identification}. He shows that both the endogenous and exogenous interaction effects can be identified if there is sufficient variation in group sizes. However, with large groups,  identification can be weak in the sense that the estimator converges in distribution  at low rates (\cite{lee2007identification}). The low rate of convergence means that we need a larger sample to have enough exogenous variation. Indeed as the group size increase, the marginal effect of an individual on its peer becomes small an more observations are needed for identification.

In a more general framework, \cite{bramoulle2009identification} investigate identification and estimation of network effect. They use the structure of the
network to identify the network effect. Their identification strategy relies on the use
of spatial lags of friends' (or friends of friends') characteristics as instruments.
But, if the network is highly transitive (i.e. if a friend of my friend is also likely to be my friend),
the identification is also weak. Weak identification can also occurs if the there are too many isolated individuals, the weak identification correspond to the weak instruments as in \cite{Staigerandstock97}. This paper focuses its attention on highly transitive networks.

More recently, Liu and Lee (2010) have considered the estimation of a social network where the endogenous effect is given by the aggregate choices of an agent's friends. They show that different positions
of agents in a network captured by the \cite{bonacich1987power} centrality measure can be used
as additional instrumental variables to improve estimation efficiency.
The number of such instruments depends on the number of groups, and can be very large.
Liu and Lee (2010) propose two-stage least squares (2SLS) and generalized method of moments (GMM) estimators. The proposed estimators have an
asymptotic bias due to the presence of many instruments.

The existing papers in the literature use instrumental variable (IV) methods or quasi-maximum likelihood method to estimate the network effect. The present paper is interested on the use of IV when identification is weak in the sense describe above. We will show that, in the estimation of peer effects using IV methods, highly transitive network or large group size imply the use of highly correlated instruments (where the set of instrumental variables contains the included and excluded instruments). If the \cite{bonacich1987power} centrality measure are used the number of instruments increase with the number of groups. In both cases, the structure of the interaction generates a weak identification issue. The weak identification problem comes from the near-perfect collinearity of the first-stage regression.

This paper proposes simple-to-check necessary conditions for  identification based on the spectral decomposition of the network adjacency matrix. It shows that identification of the network effects is possible many cases. However, given that all exogenous variations come from the system, weak identification may be a concern. I propose a regularized 2SLS
estimators for network models with spatial autoregressive (SAR) representations. High-dimensional reduction
techniques are used to mitigate the finite sample bias of the 2SLS estimators
stemming from the use of many or highly correlated instruments.
The regularized 2SLS  estimators are based on three ways of computing a regularized
 inverse of the (possibly infinite dimensional) covariance matrix
of the instruments.   The regularization methods come from the literature on inverse
problems (see Kress (1999) and Carrasco, Florens, and Renault (2007)).
The first estimator is based on Tikhonov (ridge) regularization. The Tikhonov (ridge) regularization is known in the machine learning
literature for its ability to address near-perfect collinearity problems. The
second estimator is based on the iterative Landweber-Fridman method. It has the same regularization
properties as the ridge method, with the advantage of being appropriate for larger-scale problems. The third estimator is based on the principal components associated with
the largest eigenvalues. The use of the principal components is very popular for estimating models with factors. In the presence of many instruments, the use of few principal components
can help reduce the first-stage regression dimension. The regularized estimator presented in the paper depends on a tuning parameter, I proposed a data-driven method for it selection based the estimation of an approximation of the mean square error of the estimator.

The regularized 2SLS   estimators are consistent and asymptotically normal and unbiased.
The regularized 2SLS estimators achieve the semiparametric efficiency bound. However, the consistency and asymptotic normality conditions require more regularization than in Carrasco (2012).
A Monte Carlo experiment shows that the  regularized estimators perform well.
In general, the quality of the regularized estimators improves as the density of the network increases.

I demonstrate the empirical relevance of my estimators by estimating a model of tax competition between municipalities in Finland. The size of the tax competition parameter seems larger than what is suggested by \cite{lyytikainen2012tax}. However, the regularized estimators are not statistically different from zero. This leaves the conclusion unchanged that tax competition is absent between municipalities in Finland.\footnote{The inference carried out in the empirical example does not account for the effect of regularization.}

The large existing literature on network models focuses on two main issues: identification and the estimation of the network effect. In his seminal work, \cite{manski1993identification}
shows that linear-in-means specifications suffer from the reflection problem, so endogenous and contextual effects cannot be separately identified. \cite{lee2007identification} and \cite{bramoulle2009identification} propose identification strategies for a local-average network model based on differences in group sizes and structures. Liu and Lee (2010) show that the Bonacich (1987) centrality measure can also be used as
additional instruments to improve identification and estimation efficiency.  \cite{lee2007identification} and \cite{bramoulle2009identification}
use the instrumental variables method to estimate the parameter of interest.
Liu and Lee (2010) propose a generalized method of moments (GMM) estimation approach, following Kelejian and Prucha (1998, 1999), who propose 2SLS and GMM approaches for estimating SAR models. The inclusion of the measure of centrality implies the use of many moment conditions (see \cite{DonaldandNewey01}, \cite{HansenHausmanandN08} and \cite{hasselt2010many} for some recent developments in this area).

In this paper, I assume that there are many instruments at hand (they are generated by the structure imposed on the data), and therefore use a framework that allows for an infinite number of instruments. Thus, this paper contributes to the literature on models for which the number of instruments exceeds the sample size. In a linear model framework without network effects, \cite{Carrasco12} proposes an
estimation procedure that allows for the use of many instruments; the number of instruments may
be smaller or larger than the sample size, or even infinite. Moreover, \cite{carrasco2016efficient}
show that these methods can be used to improve identification in weak instrumental variables estimation.
Closely related papers also include \cite{Kuer12}, who considers a kernel-weighted
GMM estimator; \cite{okui04}, who uses shrinkage with many instruments; and \cite{BaiNg10} and \cite{KapMar10}, who also assume that the endogenous regressors depend on a small number of factors that are exogenous. Using estimated factors as instruments, they assume that the number of variables from which the factors are estimated can be larger than the sample size.
 \cite{belloni2012sparse} propose an instrumental variables estimator under the first-stage sparsity assumption.
\cite{HansenKozbur} propose a ridge-regularized jackknife instrumental variable estimator in the presence of heteroscedasticity, which does not require sparsity, and with good sizes.

Another important focus in the instrumental variables estimation literature is on weak instruments or weak identification
(see, for example, \cite{chaoandswanson05} and \cite{newey2009generalized}). In this paper, I assume that the concentration parameter grows at the same rate as the sample size. However, I allow for the possibility of weak identification resulting from near-perfect collinearity in the set of instruments. My framework is similar to \cite{caner2012cue}, with the difference that the near-singular design does not come for the proliferation of instruments, but from the structure of the social or spatial interaction.

The paper is organized as follows. Section 2 presents
the network model. Section 3 discusses identification and estimation in network models. It proposes the regularized 2SLS
approach to estimating the model. The selection of the regularization parameter is discussed in Section 4.
Monte Carlo evidence on the performance of the
proposed estimators for small samples is given in Section 5. An empirical application on local government tax competition is proposed in Section 6. Section 7 concludes.

\section{ The Model}

The following social interaction model is considered:
\begin{equation}\label{Modl}
    Y_r= \lambda W_r Y_r + X_{1r}\beta_1+ W_rX_{2r}\beta_2+ \iota_{m_r}\gamma_r+ u_r
\end{equation}
 with $u_r=\rho M_r u_r+\varepsilon_r$ and $r=1...\bar{r}$,  where $\bar{r}$
 is the total number of groups and $m_r$ is the number of individuals in group $r$.

 $Y_r=(y_{1r},...,y_{m_rr})'$ is an $m_r$-dimensional vector that represents the outcomes of interest. $y_{ir}$ is the
 observation of individual $i$ in group $r$.
 The total number of individuals in the sample is $n=\sum_{r=1}^{\bar{r}} m_r$.

 $W_r$ and $M_r$ are $m_r \times m_r$ sociomatrices of known constants, and may or may not be the same.

 $\lambda$ is a scalar that captures endogenous network effects. I assume that this effect is the same for all individuals and groups. The outcomes of individuals influences those of their successors in the network graph (the successors are usually a friends or peers).

 In such a linear model, the parameter $\lambda$ is usually interpreted as the partial effect of a one-unit change in the explanatory variable on the outcome. The explanatory variable in the present case is a product of the a sociomatice $W_r$ and friends' outcomes $Y_r$. If the sociomatrix $W_r$ is row-normalized, the endogenous network effect captured by $\lambda$ represents the expected change in the outcome of an individual if all his friends' outcomes were changed by one unit. This corresponds to the ``local average'' endogenous effect in the terminology of \cite{liu2014endogenous}. On the other hand, if $W_r$ is not row-normalized, it is impossible to know which intervention is the source of the exogenous change in $W_rY_r$ (see \cite{goldsmith2013social} and \cite{angrist2014perils} for a discussion on the causal interpretation of the network effect). The unit variation in $W_rY_r$ could come from a change in the allocation of friend, an intervention on friends' outcomes or both. This should be done in a specific manner to obtain a unit change. Such a situation corresponds to the ``local aggregate'' endogenous effect in the terminology of \cite{liu2014endogenous}.

My model specification allows for the use of the ``local average'' and ``local aggregate'' endogenous effects. Micro-foundations developed in \cite{liu2014endogenous} suggest that ``local average'' should be used in situations where the network effect comes from individuals trying to conform to the social norm and the ``local aggregate'' for a situation where there is leakage.

 $X_{1r}$ and $X_{2r}$ are $m_r \times k_1$ and $m_r \times k_2$ matrices, respectively. They represent individuals' exogenous characteristics.
 $\beta_1$ is the parameter measuring  the
dependence of individuals' outcomes on their own characteristics.
The outcomes of individuals may also depend
on the characteristics of their predecessors via the exogenous
contextual effect, $\beta_2$. $\iota_{m_r}$ is an $m_r$-dimensional vector of ones and
$\gamma_r$ represents the unobserved group-specific effect (it is treated as a vector of unknown parameters that will not be estimated).

Aside from
the group fixed effect, $\rho$ captures unobservable correlated effects
between individuals and their connections in the network.

 $\varepsilon_r$ is the $m_r$-dimensional disturbance vector,
 $\varepsilon_{ir}$ are $i.i.d.$ with a mean of 0 and variance of $\sigma^2$ for all $i$ and $r$.
I define $X_r= (X_{1r}, W_rX_{2r})$.

 For a sample with $\bar{r}$ groups,
the data is stacked up by defining  $V=(V_1',...,V_{\bar{r}}')'$ for $V=Y,X,\varepsilon$ or $u$.

I also define $W=D(W_1, W_2,...,W_{\bar{r}})$ and $M=D(M_1, M_2,...,M_{\bar{r}})$, $\iota=D(\iota_{m_1}, \iota_{m_2},...,\iota_{m_{\bar{r}}})$,
where $ D(A_1,.., A_K)$ is a block diagonal matrix in which the diagonal blocks are $m_k\times n_k$ matrices, denoted as $A_k$, for $k=1,...,K$.

 The full sample model is
 \begin{equation}
   Y=\lambda W Y + X \beta + \iota \gamma +u
   \label{Model}
   \end{equation}
  where $u=\rho M u+ \varepsilon$.

  I define $R(\rho)=(I-\rho M)$. The Cochrane-Orcutt-type transformation of the model is obtained  by multiplying equation (\ref{Model}) by $R=R(\rho_0)$, where
  $\rho_0$ is the true value of the parameter $\rho$:  $$RY=\lambda R W Y + RX \beta + R\iota \gamma  +Ru.$$
  This leads to the following equation:
  \begin{equation}
     RY=\lambda R W Y + RX \beta + R\iota \gamma  +\varepsilon.
     \label{Model2}
     \end{equation}

When the number of groups is large, we have the incidental parameter problem (see \cite{NeyManScott1948} and \cite{lancaster2000incidental} for a discussion of the consequences of this problem).

To eliminate unobserved group heterogeneity, I define $$J_r=I_{m_r}-(\iota_{m_r},M_r\iota_{m_r}) [(\iota_{m_r},M_r\iota_{m_r})'(\iota_{m_r},M_r\iota_{m_r})]^{-}(\iota_{m_r},M_r\iota_{m_r})'$$
where $A^-$ is the generalized inverse of a square matrix $A$.  In general, $J_r$ represents the projection of  an $m_r$-dimensional vector
	 on the space spanned by $\iota_{m_r}$ and $M_r\iota_{m_r}$ if  they are linearly independent.
Otherwise, $J_r=I_{m_r}-\frac{1}{m_r}\iota_{m_r}\iota_{m_r}'$, which is the deviation from the group mean projector.

The matrix $J=D(J_1, J_2,...,J_{\bar{r}})$ is then pre-multiplied by equation (\ref{Model2}) to create a model without the unobserved group-effect parameters:
  \begin{equation}
       JRY=\lambda JR W Y + JRX \beta + J\varepsilon.
       \label{Model3}
     \end{equation}
This is the structural equation, and we are interested in the estimation of $\lambda$, $\beta_1$, $\beta_2$ and $\rho$. The discussions on the identification and estimation of $\lambda$, $\beta_1$ and $\beta_2$ in this paper will be carried out under the assumption of a consistent estimation of $\rho$.

I define $S(\lambda)=I-\lambda W$. I assume that equation (\ref{Model}) is an equilibrium and that $S \equiv S(\lambda_0)$ is invertible at the true parameter value. The
   equilibrium vector $Y$ is given by the reduced-form equation:
     \begin{equation}
        Y= S^{-1}(X \beta + \iota\gamma )+  S^{-1}R^{-1}\varepsilon.
        \label{Model4}
     \end{equation}

     It follows that $WY= WS^{-1}(X \beta + \iota \gamma )+  WS^{-1}R^{-1}\varepsilon $ and $WY$ is correlated with $\varepsilon$.
     Hence, in general, equation (\ref{Model3}) cannot be consistently estimated by ordinary least squares (OLS). Moreover, this model may not be considered as a self-contained
     system where the transformed variable $JRY$ can be expressed as a function of the exogenous variables and disturbances. Hence, a partial-likelihood-type approach based only on equation (\ref{Model3}) may not be feasible.

In this paper, I consider the estimation of the parameters of equation ($\ref{Model3})$ using regularized 2SLS.\footnote{An extension would be to estimate the same model using a limited information maximum likelihood (LIML) method (the least variance ratio (LVR)) from \cite{carrasco2015regularized}. In models with independent observations, the LIML estimator can also be derived using the LVR principle (\cite{davidsonMackinnon93}). The LVR estimator is not equivalent to the LIML estimator for the SAR model. This is analogous to the difference between the 2SLS and maximum likelihood estimators for the SAR model (\cite{lee2004asymptotic}).}

\section{Identification and Estimation of the Network Models}
This section presents the identification and estimation of the network model parameters using regularization techniques. It first discusses weak identification in network model.   It, then, proposes a regularized 2SLS model using three regularized methods (Tikhonov, Landweber-Fridman and principal component). They are presented in a unified framework covering both a finite or infinite number of instruments. The focus is on estimating endogenous and contextual effects under the assumption of a preliminary estimator of the unobservable correlation between individuals and their connections in the network. I also derive the asymptotic properties of the models' estimated parameters.\\

\subsection{Identification and Network Structure}

The model presented in  Equation(\ref{Model}) proposes an underlying structure assumed to have generated the data of the population from which our sample is drawn. The estimation strategy that I  propose later aims at making statements about the parameters of this model.  To that end, they shouldn't exist many parametrizations compatible with the observed data. Discussing conditions under which a unique parametric characterisation exist is a considerable problem in the estimation of network models (see \cite{bramoulle2009identification})  and, in econometrics (see \cite{dufour2010identification} for a general discussion on identification).

The discussion on the identification is done under a number of assumptions.

 \textbf{ Assumption 1.} The elements of $\varepsilon_{ir}$ are $i.i.d.$ with a mean of 0 and variance of $\sigma^2$, and a moment of order higher than the fourth exists.

\textbf{Assumption 2. }The sequences of matrices $\{W\}$, $\{M\}$, $\{S^{-1}\}$ and
 $\{R^{-1}\}$ are uniformly bounded (UB), and $Sup \|\lambda W \|<1$.\footnote{Uniformly bounded in row (column) sums of the absolute value of a sequence of square matrices $\{A\}$ will be abbreviated as
 UBR (UBC), and uniformly bounded in both row and column sums in absolute value as UB. A sequence of square matrices $\{A\}$, where $A = [A_{ij}]$, is said to be UBR (UBC) if the sequence of the row-sum matrix norm of $A$ (column-sum matrix norm of $A$) is bounded.}\newline

I take $\varepsilon(\rho_0, \delta) = JR(Y - Z\delta )=f( \delta_0- \delta)+JRWS^{-1}R^{-1}\varepsilon(\lambda_0-\lambda)+ J\varepsilon$, with $f=JR[WS^{-1}(X\beta_0+\iota \gamma_0), X]$, where $\lambda_0$, $\beta_0$ and  $\gamma_0$ are true values of the parameters $\delta=(\lambda, \beta')'$ and $Z=( WY, X )$.

Under Assumption 2 (i.e. that $Sup \|\lambda W \|<1$), $f$ can be approximated by a linear combination of $(WX, W^2X, . . .)$ ,
$(W\iota, W^2\iota, . . .)$ and $X$. This is a typical case where the number of potential instruments is infinite.

I define $Q=J[Q_0,  MQ_0]$, where $Q_0=[WX, W^2X, . . .W\iota, W^2\iota, . . ., X]$ is the infinite dimensional set of instruments.

We can also consider the case where only a finite number of  instruments, such as $m_1<n$, is used.\footnote{There may be a finite set of instruments when the network effect is very small, such that $\lambda^m \rightarrow 0$ as $m\rightarrow \infty$ at a very fast rate.}
For this case, I define $$Q_{m_1}=J[Q_{0m_1}, MQ_{0m_1}] $$ where $Q_{0m_1}=[WX, W^2X, . . .W^{m_1}X, W\iota, W^2\iota, . . .,W^{m_1}\iota, X]$.

As discussed in \cite{LeeandLung2010Central}, $\delta$ is identified if $Q_{m_1}'f$
has full column rank $k+1$. This rank condition requires that $f$ has full rank $k+1$. Note that this assumes that $Q_{m_1}$ is full column rank (meaning no perfect collinearity between instruments). If instruments are near-perfectly or perfectly collinear, $f$ having full rank $k+1$ does not ensure identification.\footnote{Section 3.1 discusses  the effect of near-perfect collinearity on the identification of the network effect.}
If $W_r$ does not have equal  degrees in all its nodes and $W_r$ is not row-normalized, the centrality score of each individual in his group helps to identify $\delta$.
This is possible even if $\beta_0=0$. However, if $W_r$ has constant row sums, then $f = JR[WS^{-1}X\beta_0, X]$ and the
identification is impossible for  $\beta_0=0$. Under Assumptions 1 and 2, $\delta$ is identified.\footnote{These identification results are  from \cite{LeeandLung2010Central}. My work generalizes the results to an infinite number of instruments.}

The identification in the general case with an infinite number of instruments is possible if the matrix with an infinite number of rows, $Q'f$, has full column rank. The identification is based on the moment condition $E(Q'\varepsilon(\rho_0, \delta))=0$ (i.e. $Q'f(\delta_0-\delta)=0$).

For any sample size $n$, $rank(Q) \leq n$. If we assume that $rank (QQ')=n$,  then the full column rank condition only requires that $f$ has full rank $k+1$.\footnote{ Section 3.2 proposes regularization tools that can ensure the identification of $\delta$ with a regularized version of the orthogonality condition.}
The same identification conditions as in the finite dimensional case follow.

The identification of the model parameters relies on the structure of the network through the adjacency matrix $W$.   The adjacency matrix is an $n \times n$ matrix. Let  $\tau_1\geq \tau_2\geq...\geq \tau_n$ be its $n$ eigenvalues. An eigenvalue could have multiplicity one or $k$ depending on the number of corresponding eigenvectors. Let define $\varrho_w$, to be the number of distinct eigenvalues of the adjacency matrix. The results propose in proposition 1 to 3  apply to symmetric spatial and adjacency matrix $W$.  Undirected networks' adjacency matrices is an example of a network structure represented by a symmetric adjacency matrix.

{\begin{prop}
Consider a network model represented by Equation \ref{Model} with $\rho=0$. If  $\varrho_w=2$, then the network
effects are not identified.
\end{prop}}

Proposition 1 implies that the identification of the network effect can be reduced to a spectral analysis of the adjacency matrix.  It provide a easy to verify condition for identification of the network effects under the assumption of network exogeneity. Indeed, if $\varrho_w=2$, using the Cayley-Hamilton theorem, I can show that there exist $\mu_0$ and $\mu_1$ non null scalars such that $W^2= \mu_0I+\mu_1W$. Then, using proposition 1 from \cite{bramoulle2009identification}  the network effects are not identified.

{\begin{prop}
Consider a network model represented by Equation \ref{Model} with $\rho_0=0$ and $\varepsilon(\delta) = J(Y - Z\delta )=f( \delta_0- \delta)+JWS^{-1}\varepsilon(\lambda_0-\lambda)+ J\varepsilon$, with $f=J[WS^{-1}(X\beta_0+\iota \gamma_0), X]$, where $\lambda_0$ and $\beta_0\neq0$ are true values of the parameters and $\delta=(\lambda, \beta')'$ and Assumptions 1 and 2 hold. Let $\varrho_w$ be the number of distinct eigenvalues of the adjacency matrix W.
If $[WX, W^2X, . . ., W^{\varrho_w-1}X, X ]$  is full rank column, the network effects are identified.

\end{prop}}

Proposition 2 gives a relationship between the identification of network effect and the spectral decomposition of the adjacency matrix. If $\varrho_w=2$ using the definition of $X$ and applying the Cayley-Hamilton theorem leads to the conclusion that $[WX, X]$ is not full rank column. Thus, $JWX$ cannot be excluded from the structural equation and, therefore, can not sere as an instrumental variable for $JWY$.   However, if the number of distinct eigenvalues is strictly greater than 2, identification may be possible. For instance if $\varrho_w=3$, $\rho_0=0$  and $[WX, W^2X, X ]$ is full rank column  then the network effect are identified. Indeed, $JWX$ and $ JW^2X$  serve as excluded instruments for $JWY$.

The full rank condition can be generalised to a necessary and sufficient condition, under very restrictive assumptions on the set in which the true model's parameters belong. This possibility is discussed in the proof of proposition two in the appendix.

I now consider the case where there is spatial serial correlation. The following proposition generalizes proposition 1 and 2.
{\begin{prop}
Consider a network model represented by Equation \ref{Model}, $\beta_0\neq 0$ and Assumptions 1 and 2 hold. Let $\varrho_w$ be the number of distinct eigenvalues of the adjacency matrix $W$. If $Q_{\varrho_w}=[Q_{0\varrho_w}, MQ_{0\varrho_w}] $ where $Q_{0\varrho_w}=[WX, W^2X, . . .W^{\varrho_w-1}X, W\iota, W^2\iota, . . .,W^{\varrho_w-1}\iota, X]$  is full rank column, the network effects are identified.

\end{prop}}

A special case of a model with spatial serial correlation is one in which $W=M$. In such a situation, proposition 3 becomes similar to Proposition 2.  Otherwise, the identification of the network effects could be achieved via the effect of unobserved shock on peers of peers via $M$. Having spatial correlation provides a second source of exogenous variation.

The identification of the network effects seems to rest upon the possibility of having a full rank column matrix $Q_{\varrho_w}=J[WX, W^2X, ... , W^{\varrho_w-1}X, X ]$.  The rank property of $Q_{\varrho_w}$  can be measured by condition number of the matrix $Q_{\varrho_w}Q_{\varrho_w}'$.\footnote{The condition number is the ratio between the largest and the smallest eigenvalue of a symmetric matrix (see \cite{ozturk2000ill} for the relation between ill-conditioned and multicollinearity).} Large values of the condition number correspond to a situation of near-rank deficiency and near-non-identification of the models network effects. I consider a model with near rank deficient $Q_{\varrho_w}$  matrix as being weakly identified following the terminology of \cite{dufour2010identification}. The following subsection provides a discussion of the empirical contexts in which existing network effects identification strategy may become weak.

\subsection{Weak Identification in Network Models}

Since Manski (1993), the identification problem in network models has been a major concern for econometricians.
After finding that separately identifying endogenous and exogenous interaction effects in a linear-in-mean model is not possible, many subsequent studies have investigated network structures in which identification is possible.
The identification of the network effect is achieved through group size variation or by exploiting the structure of the network. It is notable that in all cases, additional information is required to overcome the reflection problem.

Lee (2007) uses variations in group sizes to identify both the endogenous and exogenous interaction effects. His identification relies on having sufficient variation in group size. For example, if we assume that the we have two groups form $m_1$ and $m_2$ individual and we consider the adjacency matrix formed as follows $W_{ii}=0$ and $W_{ij}=\frac{1}{m_k-1}$ as long as $i$ and $j$ belong in to the same group $k$.  $W$ can be represented as a block diagonal matrix. Its distinct eigenvalues are $\tau_{1}= -\frac{1}{m_1-1}$, $\tau_{2}= -\frac{1}{m_2-1}$ and $\tau_{3}=  1$.
If the group sizes are equal, we have exactly two distinct eigenvalues. An the network effect cannot be identified. Different group sizes lead to more than two distinct eigenvalues. The spectral decompossition of the adjacency matrix leads to the same conclusion as in the comments from \cite{bramoulle2009identification} on Lee's identification with two groups of different same sizes. I can show that with group  large group size there is almost near-perfect collinearity between $WX, W^2X, . . ., W^{\varrho_w-1}X$ and, $X$.  Or  in other words, with large groups, the identification can be weak.

More precisely, let us consider the model presented in Section 2. To focus the discussion on the possibility of model's weak identification, we will consider the version of the social interaction model without spatial serial correlation.

For an individual in group $r$, the model above gives
\begin{equation}\label{Lee}
y_{ir}=\lambda \left( \frac{1}{m_r-1} \sum_{j\neq i}^{m_r} y_{jr}\right) + x_{1ir} \beta_1 + \left(\frac{1}{m_r-1} \sum_{j\neq i}^{m_r} x_{2jr} \right) \beta_2 + \gamma_r +\varepsilon_{ir}.
\end{equation}

The reduced form after a within transformation is given by:
\begin{equation}\label{reduceLee}
y_{ir}-\bar{y}_r= (x_{1ir}-\bar{x}_{1r})\frac{(m_r-1)\beta_1}{m_r-1+\lambda} -(x_{2ir}-\bar{x}_{2r})\frac{\beta_2 }{m_r-1+\lambda}  + \frac{m_r-1 }{m_r-1+\lambda}(\varepsilon_{ir}-\bar{\varepsilon}_r)
\end{equation}
where $\bar{y}_r$, $\bar{x}_{1r}$, $\bar{x}_{2r}$, and $\bar{\varepsilon}_r$  are the group average of the variables excluding individual $i$ (see equation 12 in \cite{bramoulle2009identification}, and equation 2.5 in Lee (2007)). To simplify the discussion, without loose of generality, let assume that $x_{1ir}=x_{2ir}$. Thus,
\begin{equation}\label{reduceLee1}
y_{ir}-\bar{y}_r= (x_{1ir}-\bar{x}_{1r})\frac{(m_r-1)\beta_1-\beta_2}{m_r-1+\lambda} + \frac{m_r-1 }{m_r-1+\lambda}(\varepsilon_{ir}-\bar{\varepsilon}_r)
\end{equation}

Each reduced form equation gives value for $\frac{(m_r-1)\beta_1-\beta_2}{m_r-1+\lambda}$. Identification of the parameters in this model comes from the variations in $\frac{(m_r-1)\beta_1-\beta_2}{m_r-1+\lambda}$. Indeed, \cite{bramoulle2009identification} show that we need at least three different group sizes to be able to identify $\beta_1$, $\beta_2$ and, $\nu$. The parameters are obtained after solving a system of linear  equations. There is a need for at least three distinct equations for a unique solution.

When the group size becomes large, $\frac{(m_r-1)\beta_1-\beta_2}{m_r-1+\lambda}$  converge to  constants, which means no or very small variation as $m$ becomes large in the coefficient of the reduced form of all groups. A explanation is that with a large group, the marginal contribution of an additional member of the group is relatively small, which means that  the amount of exogenous variation useful for identification vanishes as the group's size increases. This situation is a case of weak identification of the network effects.

The spectral decomposition of the adjacency  matrix associated with Lee's model is a block diagonal matrix. The distinct eigenvalues are given by $\tau_r=-\frac{1}{m_r-1}$ $r=1,...,\bar{r}$ and $\tau_{\bar{r}+1}=1$. As the group sizes increase, the difference between the eigenvalue $\tau_r$ decreases. The number of distinct eigenvalue because nearly equal to two and the model is weakly identified. It can then be said that if the groups are large based on Proposition 1, $X$, $WX$ and $W^2X$ will be nearly linearly dependent, leading to weak identification.

\cite{bramoulle2009identification} use the structure of the network to identify the network effect. Their work proposes a general framework that incorporates Lee's and Manski's setups as special cases.
The identification strategy proposed in their work relies on the use
of spatial lags of friends' (i.e. friends of friends') characteristics as instruments.
The variables $WX, W^2X$ and $ W^3X...$ are used as instruments for $WY$.
The condition for identification is that $I, W$ and $W^2$ (or, as noted in Proposition 1 and 4 of \cite{bramoulle2009identification}, $I, W, W^2$ and $ W^3$ in the presence of correlated effects) are linearly independent.
Variation in group size ensures  that $I, W$ and $ W^2$ are linearly independent.
 If the network is highly transitive (i.e. a friend of my friend is likely to be my friend too; $W\sim W^2$), identification is also weak. In practice, using  $WX, W^2X$ and $ W^3X...$ as instruments can lead to near-perfect collinearity, which implies weak identification (\cite{gibbons2012mostly}).\footnote{The extreme case of fully connected graph has exactly two distinct eigenvalues. An the application of the Proposition 1 implies that the network effects are not identified.}
The the nearly violation of the full rank condition of Proposition 2 is a potential source of  weak identification. Because, it leads to a near-perfect collinearity occurring in the first-stage regression of the endogenous network effect. The use of regularization methods, such as ridge regression, has been shown to solve these problems.

Liu and Lee (2010) also consider the estimation of a social interaction network model.
As in \cite{bramoulle2009identification}, they exploit the structure of the network to identify the network effect. In addition to $WX, W^2X$ and $ W^3X...$, the Bonacich centrality across nodes in a network is used as an instrumental variable to identify network effects and improve estimation efficiency.  The use of the Bonacich centrality measure usually leads to the use of many instruments.
The 2SLS estimates obtained with these instruments are biased because of the large number of instruments used. Liu and Lee (2010) propose a bias-corrected 2SLS method to account for this.

In this paper, I use regularization techniques. These high-dimensional  estimation techniques enable the use of all instruments and deliver efficiency with better finite sample properties (see \cite{Carrasco12} and \cite{carrasco2015regularized}).
 In this case, asymptotic efficiency can be obtained by using many (or all potential) instruments. I use both the Bonacich centrality measure and $WX, W^2X$ and $ W^3X...$ as instrumental variables and apply a high-dimensional technique to mitigate the problem of near-perfect collinearity resulting from network structure or/and the bias of many instruments.

\subsection{Estimation Using Regularization Methods}

The parameters of interest can be estimated using instrumental variables. We can use a finite number of instruments or all potential instruments. As the number of instruments increases, estimation becomes  asymptotically more efficient. However, a large number of instruments relative to the sample size creates the many instruments problem (see, for example, \cite{Bekker94}, \cite{DonaldandNewey01} and \cite{hanphillips2006}). The parameter of interest can also be weakly identified when a fixed number of instruments is used but the structure of the interaction does not provide sufficient exogenous variation. In such cases, using a fixed number of instrumental variables will not avoid the bias problem in the estimation.

The 2SLS estimator with a fixed number of instrumental variables will be consistent and asymptotically normal, but may be less efficient than using many instruments.
In order to use all potential instruments ($Q$), I use regularization tools.
In addition to addressing the many instruments bias in \cite{Carrasco12}, my objective is to use regularization to address the weak identification problem.

Let $\varepsilon(\rho_0, \delta) = JR(Y - Z\delta )$, with $\delta=(\lambda, \beta')'$ and $Z=( WY, X )$.
 The estimation is based on moments corresponding to the orthogonality condition of $Q$ and $J\varepsilon$ given by\footnote{The set of instrumental variables is $Q=J[Q_0,  MQ_0]$ with $Q_0=[WX, W^2X, . . .W\iota, W^2\iota, . . ., X]$. They can be normalized or standardized.}
 \begin{eqnarray}\label{momentsar}
 E(Q'\varepsilon(\rho_0, \delta))=0
 \end{eqnarray}

My identification results are conditional on $\rho_0$. I should first have a preliminary estimator
$\tilde{\rho}$ of $\rho$. I take $\tilde{R}= I-\tilde{\rho}M$ to be an estimator of $R$.

The regularized estimators used in this paper require the definition of some mathematical objects. My notation follows existing notation in the literature on regularization methods. The set of all potential instrumental variables ($Q$) is a countable infinite set. $\pi $ is a positive measure on $\mathbb{N}$,\footnote{For a detailed discussion on the role and choice of $\pi $, see \cite{Carrasco12} and \cite{carrascoandflorens2004} when $\pi$ is a measure on $\mathbb{R}$. In my model, $\pi$ can, for example, be $\pi_k=\frac{\lambda^k}{\sum_{k \in \mathbb{N}} \lambda^k }$ with $k \in \mathbb{N}$. } and $l^{2}(\pi )$ is the Hilbert space
of square-summable sequences with respect to $\pi $ in the real space. I define the
covariance operator $K$ of the instruments as
\begin{equation*}
K:l^{2}(\pi )\hspace{0.03in}\rightarrow \hspace{0.03in}l^{2}(\pi )
\end{equation*}%
\begin{equation*}
(Kg)_j=\sum_{k \in \mathbb{N} } E(Q_{ji}Q_{ki} g_k\pi_k)
\end{equation*}%
where $Q_{ji}$ is the $j^{th}$ column and $i^{th}$ line of $Q$. Under the assumption that $|Q_{ji}Q_{ki}|$ for all $j,k$ and $i$ are uniformly bounded, $K$ is a compact operator (see \cite{carrascoandFlorens2007} for a definition). Indeed, under Assumption 2,  the operator $K$ is a   Hilbert-Schmidt
operator; I assume that it has non-zero eigenvalues.\footnote{I assume that the element of $X$ are uniformly bounded.}

 I consider $\nu _{j};\hspace{0.05in}j=1,2,...$ to be the eigenvalues (in decreasing order) of $K$, and $\hspace{0.05in}\phi _{j};\hspace{0.05in}j=1,2,...$ to be the orthogonal eigenvector of $K$. $K$ can be estimated by $K_{n}$, defined as:

\begin{equation*}
K_n:l^2(\pi)\hspace{0.03in}\rightarrow \hspace{0.03in}l^2(\pi)
\end{equation*}
\begin{equation*}
(K_n g)_j= \sum_{k \in \mathbb{N} } \frac{1}{n}\sum^{n}_{i=1}Q_{ji}Q_{ki}g_k\pi_k.
\end{equation*}

In the SAR model, the number of potential moment conditions  can be infinite as in equation (\ref{momentsar}).  Therefore, the inverse of $K_{n}$ needs
to be regularized because it is nearly singular. By definition (see \cite{Kress99}, p. 269), a regularized inverse of an operator $K$ is
\begin{equation*}
R_{\alpha }:l^{2}(\pi )\hspace{0.03in}\rightarrow \hspace{0.03in}l^{2}(\pi )
\end{equation*}%
such that $\lim_{\alpha \rightarrow 0}R_{\alpha }K\varphi =\varphi $, $%
\forall \hspace{0.02in}\varphi \in l^{2}(\pi ).$

I consider three different types of regularization schemes: Tikhonov (T),
Landweber-Fridman (LF) and principal component
(PC). They are defined as follows:

\begin{itemize}

\item \textbf{Tikhonov (T)}\newline
Tikhonov regularization is also known as ridge regularization:\newline
$$(K^{\alpha })^{-1}r=(K^{2}+\alpha I)^{-1}Kr$$ or
\begin{equation*}
(K^{\alpha })^{-1}r=\overset{\infty }{\underset{j=1}{\sum }}\frac{\nu
_{j}}{\nu _{j}^{2}+\alpha }\big<r,\phi _{j}\big>\phi _{j}
\end{equation*}%
\newline
where $\alpha >0$ and $I$ is the identity operator.

\item \textbf{Landweber-Fridman (LF)}

 Let $0<c<1/\Vert K\Vert ^{2}$, where $\Vert K\Vert $ is the largest eigenvalue of K (which can be estimated
by the largest eigenvalue of $K_{n}$). Then,

\begin{equation*}
(K^{\alpha })^{-1}r=\overset{\infty }{\underset{j=1}{\sum }}\frac{%
[1-(1-c\nu _{j}^{2})^{\frac{1}{\alpha }}]}{\nu _{j}}\big<r,\phi _{j}%
\big>\phi _{j}
\end{equation*}
where $\frac{1}{\alpha }$ is some positive integer.
\end{itemize}

\item \textbf{Principal component (PC)}

This method  consists of using the first eigenfunctions:
\begin{equation*}
(K^{\alpha })^{-1}r=\overset{1/\alpha }{\underset{j=1}{\sum }}\frac{1}{%
\nu_{j}}\big<r,\phi_{j}\big>\phi_{j}
\end{equation*}%
where $\frac{1}{\alpha }$ is some positive integer.\footnote{$\big<.,.\big>$ represents the scalar product in $l^{2}(\pi )$ and
in $\mathbb{R}^{n}$ (depending on the context).}
The use of PC in the first stage is equivalent to
projecting on the first principal components of the set of instrumental variables.\\

In the case of a finite number of moments, $P_{m_1}=Q_{m_1}(Q_{m_1}'Q_{m_1})^-Q_{m_1}'$ is the projection matrix on the space of instruments.
The matrix $Q_{m_1}'Q_{m_1}$ may become nearly singular when $m_1$ gets large. Moreover, when $m_1>n$, $Q_{m_1}'Q_{m_1}$ is
singular. To address these cases, I consider a regularized version of the inverse of the matrix $Q_{m_1}'Q_{m_1}$.

I use $\psi_j$ to represent the
 eigenvectors of the $n\times n$ matrix $Q_{m_1}Q_{m_1}'/n$ associated with eigenvalues, $\nu_{j}$.
 For any vector $e$, the regularized version of $P_{m_1}$, $P_{m_1}^{\alpha}$
 is:
 $$P_{m_1}^{\alpha}e=\dfrac{1}{n} \sum_{j=1}^{n}q(\alpha, \nu_j^2)\big<e,\psi_{j}\big>\psi_{j}$$
 where for T, $q(\alpha ,\nu _{j}^{2})=\frac{\nu _{j}^{2}}{\nu_{j}^{2}+\alpha }$;
 for LF, $q(\alpha ,\nu _{j}^{2})=[1-(1-c\nu _{j}^{2})^{1/\alpha }]$;
and for PC, $q(\alpha ,\nu _{j}^{2})=I(j\leq 1/\alpha )$.

The network models suggest the use of an infinite number of instruments, which is the reason we are not using instrument selection methods. Following \cite{carrascoandFlorens2000}, I define the counterpart of $P^{\alpha}$ for an infinite number of instruments as\\
$$P^{\alpha}=G(K_n^{\alpha})^{-1}G^*$$  where $G:l^{2}(\pi )\hspace{0.03in}\rightarrow \hspace{0.03in}\mathbb{R}^{n}$ with
\begin{equation*}
Gg=\left( \left \langle Q_{1},g\right \rangle ^{\prime },\left \langle
Q_{2},g\right \rangle ^{\prime },...,\left \langle Q_{n},g\right \rangle
^{\prime }\right) ^{\prime }
\end{equation*}%
and $G^{\ast }:\mathbb{R}^{n}\hspace{0.03in}\rightarrow \hspace{0.03in}
l^{2}(\pi )$ with
\begin{equation*}
G^{\ast }v=\frac{1}{n}\overset{n}{\underset{i=1}{\sum }}Q_{i}v_{i}
\end{equation*}%
such that $K_{n}=G^{\ast }G$ and $GG^{\ast }$ is an $n\times n$ matrix with a
typical element $\frac{\big<Q_{i},Q_{j}\big>}{n}$. Let $\phi_{j}$, $
\nu_{1}\geq \nu_{2}\geq ...>0$, $j=1,2,...$ be the
orthonormalized eigenvectors and eigenvalues of $K_{n}$, and $\psi _{j}$ be the eigenfunctions of $GG^{\ast }$.

 $G\phi_{j}=\sqrt{\nu _{j}}\psi _{j}$ and $G^{\ast }\psi _{j}=\sqrt{\nu _{j}}\phi_{j}$.
Note  that in this case for $e\in \mathbf{R}^{n}$, $P^{\alpha }e=\overset{\infty }%
{\underset{j=1}{\sum }}q(\alpha ,\nu _{j}^{2})\big<e,\psi _{j}\big>\psi_{j}.$
We can also note that:\begin{eqnarray}
v^{\prime }P^{\alpha }w &=&v^{\prime }G(K_{n}^{\alpha })^{-1}G^{\ast }w
\notag \\
&=&\left \langle (K_{n}^{\alpha })^{-1/2}\overset{n}{\underset{i=1}{\sum }}%
Q_{i}\left( .\right) v_{i},(K_{n}^{\alpha })^{-1/2}\frac{1}{n}\overset{n}{%
\underset{i=1}{\sum }}Q_{i}\left( .\right) w_{i}\right \rangle .  \label{PP}
\end{eqnarray}

Our objective is to estimate the parameters of the model.

I consider $S_n(k)=\dfrac{1}{n}\sum_{i=1}^n(\check{Y}_i-\check{Z}_i\delta)Q_{ik}$ with $\check{Y}=\tilde{R}Y$ and  $\check{Z}=\tilde{R}Z$.

And I denote $(K_n^{\alpha})^{-1}$ as the regularized inverse of $K_n$ and  $(K_n^{\alpha})^{-1/2}=((K_n^{\alpha})^{-1})^{1/2}$.

 The regularized 2SLS estimator of $\delta$ is defined as:
 \begin{equation}
 \hat{\delta}_{R2sls}=argmin \big<(K_n^{\alpha})^{-1/2}S_n(.),(K_n^{\alpha})^{-1/2}S_n(.)\big>.
 \label{2slseq}
 \end{equation}

 Solving the minimization problem, we have
 \begin{equation}
  \hat{\delta}_{R2sls}=(Z^{\prime }\tilde{R}^{\prime } P^{\alpha }\tilde{ R} Z)^{-1}Z^{\prime }\tilde{R}^{\prime } P^{\alpha }\tilde{R}Y.
  \label{2slsr}
 \end{equation}%
 Equation (\ref{2slsr}) defines the regularized 2SLS.  The regularized 2SLS for SAR is closely related to the regularized 2SLS of \cite{Carrasco12} and the 2SLS of Liu and Lee (2010).
 It extends \cite{Carrasco12} by considering SAR models and differs from Liu and Lee (2010) in that the  projection matrix $P$ is replaced by its regularized counterpart $P^{\alpha}$.

The 2SLS estimators proposed in this paper are for cases with spatial serial correlation and homoscedastic errors. Extending the regularization approach to deal with heteroscedasticity is left for future research. Indeed, in a companion paper, I propose regularized GMM estimators allowing the joint estimation of all parameters of the model, and the variance covariance estimator of the estimate is obtained using an approach similar to \cite{west1987simple}.



 \subsection{Consistency and Asymptotic Distributions of the Regularized 2SLS}

The following proposition shows the consistency and asymptotic normality of the
regularized 2SLS estimators. The following extra assumptions are needed.

\textbf{Assumption 3. } $H=\lim_{n\rightarrow \infty}\dfrac{1}{n}f'f $ is a finite nonsingular matrix.\newline

\textbf{Assumption 4.} (i) The elements of $X$ are uniformly bounded, $X$ has full rank $k$, $E(\varepsilon|X)=0$,
and $\lim_{n\rightarrow \infty}\dfrac{1}{n}X'X$ exists and is nonsingular.
 \newline
(ii) There is a $\omega \geq 1/2$ such that\newline
\begin{equation*}
\overset{\infty }{\underset{j=1}{\sum }}\frac{\big<%
E(Z(.,x_{i})f_{a}(x_{i})),\phi _{j}\big>^{2}}{\nu _{j}^{2\omega +1}}%
<\infty.
\end{equation*}%

 Assumption 4 (ii) ensures that regularization allows us to obtain a good asymptotic approximation of the best instrument, $f$.

 {\begin{prop}
Under Assumptions 1-4 , $\tilde{\rho}-\rho_0=O_p(1/\sqrt{n})$ and $\alpha \rightarrow 0$. Then, the T,
LF and PC estimators satisfy:
\begin{enumerate}
\item Consistency: $\hat{\delta}_{R2sls}\rightarrow \delta _{0}$ in probability as n
and $\alpha\sqrt{n} $ go to infinity.
\item Asymptotic normality: $\sqrt{n}(\hat{\delta}_{R2sls}-\delta_0)\overset{d}{\rightarrow }\mathcal{N} \left( 0,\sigma _{\varepsilon }^{2}H^{-1}\right) $
as n and $\alpha^2\sqrt{n} $ go to infinity.
\end{enumerate}
\end{prop}}

The convergence rate of  the regularized 2SLS estimators for SAR is different from those obtained without spatial correlation.
For consistency in the SAR model, $\alpha\sqrt{n}$ must go to infinity. The Carrasco (2012) regularized 2SLS estimator is consistent with a convergence rate of $n\alpha^{\frac{1}{2}}$.
Asymptotic normality is obtained if  $\alpha^2\sqrt{n}$ goes to infinity, which is also different from the   \cite{Carrasco12} asymptotic normality condition for 2SLS.
The regularization parameter $\alpha $ is allowed to go to zero slower than in  \cite{Carrasco12}  for consistency. Compared to \cite{Carrasco12}, more regularization is needed in order to achieve appropriate asymptotic behavior.
The reinforcement of these conditions is certainly due to  regularization taking into account the spatial representation of the data. \newline
If the regularization parameter is constant, the asymptotic variance will be bigger. However, asymptotically, the use of regularization should not be needed. It is therefore reasonable to have $\alpha\rightarrow 0$.

In \cite{LeeandLung2010Central}, the 2SLS estimator is biased due to the increasing number of instrumental variables.
Interestingly, the regularized 2SLS estimator for SAR models is well-centered under the assumption that  $\alpha\sqrt{n} $ goes to infinity.

The bias of the 2SLS estimator in \cite{LeeandLung2010Central} is of the form $$\sqrt{n}b_{2sls}=\sigma^2tr(P^{\alpha}RWS^{-1}R^{-1})(Z^{\prime }R P^{\alpha} R Z)^{-1}e_1.$$

Using Lemma 1 and 2 in the Appendix, I show that the 2SLS bias is of order $\sqrt{n}b_{2sls}=O_p(\dfrac{1}{\alpha \sqrt{n}})$, which goes to zero as $\alpha \sqrt{n}$ goes to infinity.  The ability to choose the regularization parameter means that we are able to control the size of $\alpha \sqrt{n}$. Therefore, selecting the appropriate regularization parameter is crucial.

The regularization methods presented  involve the use of eigenvalues and eigenvectors. The eigenvalues obtained can vary greatly because of the difference in the variance of instrumental variables in the model. For example, $W^2X$ and $W\iota$ could have different variances. To account for this difference, I use normalized instruments in the Monte Carlo simulation. We can
also standardize the instruments, which means that regularization methods will be able to account for the difference in location and scale of the instruments.
In addition, the regularized estimator  presented in this section depend on the regularization parameter, $\alpha$. The choice of this parameter is very important for the estimators' behavior in small samples. In
Section 4, I discuss the selection of the regularization parameter.

 \section{Selection of the Regularization Parameter}

This section discusses the selection of the regularized parameter for network models.
I first derive an approximation of the mean-squared error (MSE) using Nagar-type expansion. I estimate the dominant term of the MSE, and select the regularization parameter that minimizes this term.

\subsection{Approximation of the MSE}

The following proposition provides an approximation of the MSE:

\begin{prop}
\label{propmse}
 If Assumptions 1 to 4 hold, $\tilde{\rho}-\rho_0=O_p(1/\sqrt{n})$ and $n\alpha
\rightarrow \infty $ for LF-, PC- and T-regularized 2SLS estimators, then
\begin{equation*}
n(\hat{\delta}_{R2sls}-\delta _{0})(\hat{\delta}_{R2sls}-\delta _{0})^{\prime }=Q%
(\alpha )+\hat{R}(\alpha ),
\end{equation*}%
\begin{equation}
E(Q(\alpha )|X)=\sigma _{\varepsilon }^{2}H^{-1}+S(\alpha
),\label{SR2SL}
\end{equation} and
\begin{equation*}
 r(\alpha )/tr(S(\alpha ))=o_{p}(1),
\end{equation*}%
with $r(\alpha )=E(\hat{R}(\alpha)|X)$ and
\begin{equation*}
S(\alpha )=\sigma _{\varepsilon }^{2}H^{-1}\left[ \frac{f^{\prime }\left( 1-P^{\alpha }\right)
^{2}f}{n}+ \sigma _{\varepsilon }^2\frac{1}{n}\left(\sum_j q_j \right)^2e_1\iota'D'D\iota e'_1\right] H^{-1}.
\end{equation*}%
For LF and PC, $S(\alpha )=O_{p}\left(\frac{1}{n\alpha^2}+ \alpha^{\omega}\right)$ and for T, $S(\alpha )=O_{p}\left(\frac{1}{n\alpha^2}+ \alpha^{\min(\omega,2)}\right)$, with $D=JRWS^{-1}R^{-1}$ and $e_1$ is the first unit (column) vector.
\end{prop}

For the selection of $\alpha$, the relevant dominant term $S(\alpha)$ will be minimized to achieve the smallest MSE.
$S(\alpha)$ accounts for a trade-off between the bias and variance. When $\alpha$ goes to zero, the bias term increases while
the variance term  decreases. The approximation of the regularized 2SLS estimator is similar to Carrasco-regularized 2SLS. However, the expression of the MSE is more complicated because of the spatial correlation.

\subsection{Estimation of the MSE}

The aim of this subsection is to find the regularized parameter that minimizes the conditional MSE of
$\bar{\gamma} ^{\prime }\hat{\delta}_{2sls}$ for some arbitrary $k+1\times 1$ vector, $\bar{\gamma}
$. This conditional MSE is:%
\begin{eqnarray}
MSE &=&E[\bar{\gamma}^{\prime }(\hat{\delta}_{2sls}-\delta _{0})(\hat{\delta}_{2sls}-\delta
_{0})^{\prime }\bar{\gamma} |X]  \notag \\
&\sim &\bar{\gamma} ^{\prime }S(\alpha )\bar{\gamma}  \notag \\
&\equiv &S_{\bar{\gamma} }(\alpha ).  \notag
\end{eqnarray}%
$S_{\bar{\gamma} }(\alpha )$ involves the function $f$, which is unknown. We therefore
need to replace $S_{\bar{\gamma} }$ with an estimate.

Stacking the observations, the
reduced form equation can be rewritten as
\begin{equation*}
RZ=f+v.
\end{equation*}%
This expression involves $n\times (k+1)$ matrices. We can reduce the dimension
by post-multiplying by $H^{-1}\bar{\gamma} $:
\begin{equation}
RZH^{-1}\bar{\gamma} =fH^{-1}\bar{\gamma} +vH^{-1}\bar{\gamma}
\Leftrightarrow RZ_{\bar{\gamma} }=f_{\bar{\gamma} }+v_{\bar{\gamma} }  \label{Reduf}
\end{equation}%
where $v_{\bar{\gamma} i}=v_{i}^{\prime }H^{-1}\bar{\gamma} $ is a scalar.

I use $\tilde{\delta}$ to denote a preliminary estimator of $\delta$, obtained from a finite number of instruments. I use $\tilde{\rho}$ to denote a preliminary estimator of  $\rho$, obtained by the method of moments as follows:$$\tilde{\rho}=armin \tilde{g}(\rho)'\tilde{g}(\rho)$$ where
$\tilde{g}(\rho)=[M_1\tilde{\varepsilon}(\rho), M_2\tilde{\varepsilon}(\rho), M_3\tilde{\varepsilon}(\rho)]'\tilde{\varepsilon}(\rho)$,
$$M_1=JWJ-tr(JWJ)I/tr(J),$$ $$M_2=JMJ-tr(JMJ)I/tr(J),$$   $$M_3=JMWJ-tr(JMWJ)I/tr(J),$$  and $$\tilde{\varepsilon}(\rho)=JR(\rho)(Y-Z'\tilde{\delta}).$$
$\tilde{\delta}=[Z'Q_1(Q_1'Q_1)^{-1}Q_1'Z]^{-1}Z'Q_1(Q_1'Q_1)^{-1}Q_1'Y$, where $Q_1$ is a single instrument. The residual is $\hat{\varepsilon}(\rho)=JR(\tilde{\rho})(Y-Z'\tilde{\delta}).$

Let us denote $\hat{\sigma }_{\varepsilon }^{2}=\hat{\varepsilon}(\rho)'\hat{\varepsilon}(\rho)/n$, $\hat{v}_{\bar{\gamma} }=(I-P^{\tilde{\alpha}})R(\tilde{\rho})Z\tilde{H}^{-1}\bar{\gamma}$, where $\tilde{H}$ is a consistent estimate of $H$ and $\tilde{\alpha}$ is a preliminary value for $\alpha$, $\tilde{v}_{\bar{\gamma} }=(I-P^{\tilde{\alpha}})R(\tilde{\rho})Z\tilde{H}^{-1}\bar{\gamma} $ and $\hat{\sigma}_{v_{\bar{\gamma}}}^{2} =\tilde{v}_{\bar{\gamma} }'\tilde{v}_{\bar{\gamma} }/n$.

I consider the following goodness-of-fit criteria:\newline
\textbf{Mallows $C_{p}$} (\cite{Maslow73})\newline
\begin{equation*}
\hat{\varpi}^{m}(\alpha )=\frac{\hat{v}_{\bar{\gamma} }^{\prime }\hat{v}_{\bar{\gamma} }}{n}+2%
\hat{\sigma}_{v_{\bar{\gamma} }}^{2}\frac{tr(P^{\alpha })}{n}.
\end{equation*}%
\newline
\textbf{Generalized cross-validation} (\cite{Cravenwa79})\newline
\begin{equation*}
\hat{\varpi}^{cv}(\alpha )=\frac{1}{n}\frac{\hat{v}_{\bar{\gamma} }^{\prime }\hat{v}%
_{\bar{\gamma} }}{\left( 1-\frac{tr(P^{\alpha })}{n}\right) ^{2}}.
\end{equation*}%
\newline
\textbf{Leave-one-out cross-validation} (\cite{Stone74})\newline
\begin{equation*}
\hat{\varpi}^{lcv}(\alpha )=\frac{1}{n}\sum_{i=1}^{n}(\tilde{RZ}_{\bar{\gamma}_{i}}-%
\hat{f}_{\bar{\gamma}_{-i}}^{\alpha })^{2},
\end{equation*}%
\newline
where $\tilde{RZ}_{\bar{\gamma} }=W\tilde{H}^{-1}\bar{\gamma} $, $\tilde{RZ}_{\bar{\gamma}_{i}}$
is the $i^{th}$ element of $\tilde{RZ}_{\bar{\gamma} }$ and $\hat{f}_{\bar{\gamma}
_{-i}}^{\alpha }=P_{-i}^{\alpha }\tilde{RZ}_{\bar{\gamma}_{-i}}$. The $n\times
(n-1)$ matrix $P_{-i}^{\alpha }$ is such that the $P_{-i}^{\alpha
}=G(K_{n-i}^{\alpha })G_{-i}^{\ast }$ are obtained by suppressing the $i^{th}$
observation from the sample. $\tilde{RZ}_{\bar{\gamma}_{-i}}$ is the $(n-1)\times
1 $ vector constructed by suppressing the $i^{th}$ observation of $\tilde{RZ}%
_{\bar{\gamma} }$.

Using (\ref{SR2SL}), $S_{\bar{\gamma} }(\alpha )$ can be rewritten as
\begin{equation*}
S_{\bar{\gamma} }(\alpha )=\sigma _{\varepsilon }^{2}\left[ \frac{f_{\bar{\gamma} }^{\prime }\left( I-P^{\alpha
}\right) ^{2}f_{\bar{\gamma} }}{n}+ \sigma _{\varepsilon }^2\frac{1}{n}\left(\sum_j q_j \right)^2e_{1\bar{\gamma}}\iota'D'D\iota e'_{1\bar{\gamma}}\right].
\end{equation*}%

 Using \cite{Li86}'s results on $C_p$ or cross-validation procedures, note that $\hat{\varpi}(\alpha )$ approximates to

$$\varpi(\alpha)=\frac{f_{\bar{\gamma} }^{\prime }\left( I-P^{\alpha}\right) ^{2}f_{\bar{\gamma} }}{n}+\sigma_{v\bar{\gamma}}^2\frac{tr\left((P^{\alpha})^2\right)}{n}.$$

Therefore, $S_{\gamma}(\alpha)$ is estimated using the following equation:

\begin{equation*}
\hat{S}_{\bar{\gamma} }(\alpha )=\hat{\sigma} _{\varepsilon }^{2}\left[ \hat{\varpi}(\alpha)-\hat{\sigma}_{v_{\bar{\gamma} }}^{2}\frac{tr\left((P^{\alpha})^2\right)}{n}+ \hat{\sigma} _{\varepsilon }^2\frac{1}{n}\left(tr(P^{\alpha})\right)^2e_{1\bar{\gamma}}\iota'\tilde{D}'\tilde{D}\iota e'_{1\bar{\gamma}}\right]
\end{equation*}
where $\tilde{D}$ is a consistent estimator of $D$.
The optimal regularization parameter is obtained by minimising $\hat{S}_{\bar{\gamma} }(\alpha )$ with respect to $\alpha$.
My selection procedure is very similar to Carrasco (2012), and its optimality can be established using the results of \cite{Li86} and \cite{Li87}.

The regularized 2SLS process and the selection of the regularization parameters are  based on a preliminary
estimator of $\rho$. This means that if $\rho$ is not correctly estimated, the estimation of $\delta$ could be biased in an unpredictable direction. Also, the use of a cross-validation-type method to choose the regularization parameter usually influences the quality of inference. This is similar to the inference problem in non-parametric estimation (see \cite{newey1998undersmoothing} and \cite{guerre2005data}). This paper focuses on the point estimation of the parameter; post-regularization inference is left for future research.

\section{Monte Carlo Simulations}

To investigate the finite sample performance of the regularized 2SLS estimators, I conduct a simulation study based on the following model:
$$Y=\lambda_0 WY +X \beta_{10}+ WX \beta_{20}+ \iota \alpha_0+ u$$
with $u=\rho_0 M u+ \varepsilon$.

I generate four samples
with different numbers of groups ($\bar{r}$) and group sizes ($m_r$). The first
sample contains 30 groups, each with 10 individuals.
The second sample contains 60 groups, also with 10 individuals each. To study the effect of group sizes, I also consider 30 and 60 groups of 15 individuals.

For each group, the sociomatrix $W_r$ is generated as follows. First,
for the $i^{th}$ row of $W_r$ ($i = 1, . . . , m_r$), $k_{ri}$ is generated
uniformly at random from the set of integers [0, 1, 2, 3], [0, 1, ..., 6] or [0, 1, ..., 8].
Allowing for differences in the maximum number of friends helps us study the effect of the density of the network on the estimators.

The sociomatrix $W_r$ is constructed as follows. First, set the $(i + 1)th, . . . , (i + k_{ri})th$ elements of the $i^{th}$ row of $W_r$ to be 1 and the rest of the elements in that row to be 0, if $i + k_{ri} \leq m_r$. Otherwise, the entries of 1 will be wrapped around.

In the case of $k_{ri} = 0$, the $i^{th}$ row of $W_r$ will only contain zeros. $M$ is the row-normalized $W$. $X\sim \mathcal{N}(0,I)$, $ \alpha_{0r} \sim \mathcal{N}(0,0.01)$ $ \varepsilon_{r,i} \sim \mathcal{N}(0,1)$. The data are generated with $\beta_{10}=\beta_{20}=0.2$ $\lambda_0=\rho_0=0.1$.

The estimation methods considered are:
\begin{itemize}
 \item 2SLS with few instruments: the set of instruments is $Q_1=J[X ,WX, MX ,MWX]$,
\item 2SLS with many instruments: the set of instruments is $Q_2=[Q_1 ,JW\iota]$, and

\item the regularized 2SLS estimators T-2SLS (Tikhonov), LF-2SLS (Landweber-Fridman) and
PC-2SLS (principal component), with many instruments, $\tilde{Q}_2$. $\tilde{Q}_2$ is a matrix of instruments with $Q_2$'s instruments normalized to unit variance.\footnote{As noted by \cite{newey2013nonparametric}, the choice of identity for the matrix for the Tikhonov regularization method does not account for any difference in location and scale of the instruments.}

\end{itemize}

For all 2SLS estimators, a preliminary estimator of  $\rho$ is obtained by the method of moments,\\$\tilde{\rho}=argmin \tilde{g}(\rho)'\tilde{g}(\rho)$ where
$\tilde{g}(\rho)=[M_1\tilde{\varepsilon}(\rho), M_2\tilde{\varepsilon}(\rho), M_3\tilde{\varepsilon}(\rho)]'\tilde{\varepsilon}(\rho)$,\\
$$M_1=JWJ-tr(JWJ)I/tr(J),$$ $$M_2=JMJ-tr(JMJ)I/tr(J),$$   $$M_3=JMWJ-tr(JMWJ)I/tr(J),$$ $$\tilde{\varepsilon}(\rho)=JR(\rho)(Y-Z'\tilde{\delta})$$ and
$$\tilde{\delta}=[Z'Q_1(Q_1'Q_1)^{-1}Q_1'Z]^{-1}Z'Q_1(Q_1'Q_1)^{-1}Q_1'Y.$$

The selection of the regularization parameter follows the procedure proposed in Section 4. I minimise the estimated approximated MSE.

Before presenting the results of the simulations, it is important to note that the data-generating process in this experiment exhibits a very low transitivity level (there are a lot of non-connected individuals in all groups).
Moreover, the reduced-form model is sparse (for example, when the maximum number of friends is 3, $W^q=0$ for $q>4$).
The instruments coming from the relative position in the network are independent of each other. This means that high-dimensional reduction techniques should not be very effective in summarizing the information.

The simulation results, presented in Tables 2 to 7, are summarized as follows:
\begin{enumerate}
  \item The additional linear moment conditions reduce standard deviations in 2SLS estimators
of $\lambda_0$ and $\beta_{20}$. The 2SLS estimators in the model with a large number of instruments have smaller standard deviations than the 2SLS estimators in the model with a finite number of instruments.
  \item The additional instruments in $Q_2$ introduce bias into the 2SLS estimators of $\lambda_0$ and $\beta_{20}$.  The 2SLS estimators from the model with a finite number of instruments have a mean value of estimators closer to the true value of the parameter than the 2SLS estimators from the model with a large number of instruments.
  \item The regularized 2SLS  procedures substantially reduce the many instruments bias for both the 2SLS estimators, particularly in large samples.
The bias-correction estimators are similar to regularized estimators in term of bias correction for large samples from a denser network. But in small samples, the bias of the bias-corrected estimator is smaller than that of the regularized estimators. Relative to the 2SLS estimators from the model with many instruments, the regularized 2SLS estimators reduce the bias and have comparable standard deviations.

\item The performance of the regularized estimators improves with the density of the network and the number of groups. The behavior of the regularized estimator with respect to network density suggests that the regularized estimators are good candidates to improve the asymptotic behavior of the estimator of the network effect when the level of transitivity in the groups is very high.
\end{enumerate}

\section{Empirical Application: Local Tax Competition in Finland}

 The  large theoretical literature on local government tax competition can be divided in two groups: efficient local taxation (\cite{tiebout1956pure}) and tax competition models departing from Tiebout's model (\cite{lyytikainen2012tax}).  The departure from Tiebout's model leads to three types of fiscal consequences: benefit spillovers, distorting taxes on a mobile tax base, political economy considerations and information asymmetries (\cite{lyytikainen2012tax}). While the causes of local government tax interaction are certainly present in most legislation, the empirical literature has long been divided on how to identify a causal local tax competition (interaction) effect.

The identification problem here is a special case of Manski's reflection problem. In the case of municipalities in the same legislation, the network matrix can be represented by the spatial matrix of neighbors. This neighborhood structure of the municipality can be considered as exogenous with respect to tax level. I propose a model to test the hypothesis of tax competition between municipalities:

 $$T_{itr} = \lambda W_rT_{itr} + \beta_0 X_{itr}+ \beta_1 W_r X_{itr}+ \alpha_{r}+ \varepsilon_{itr}$$

 The identification and estimation of the tax competition parameter ($\lambda$) is  achieved, in a large part of the empirical literature, via two strategies.  The first strategy uses spatial lags as instruments (friends of friends' characteristics) in an instrumental variables approach, while the second uses maximum likelihood estimation, where identification is achieved via model specification. As pointed out by \cite{gibbons2012mostly}, the causality of the parameters obtained in these cases is not easy to defend. The validity of the exclusion restriction is not obvious and the correct specification of the model is not fully testable.  As an alternative, \cite{gibbons2012mostly} propose using differencing coupled with instrumental variables coming from exogenous policy variations.

 \cite{lyytikainen2012tax} estimates a tax competition parameter among Finnish local governments.  He uses changes in statutory lower limits to property tax rates as a source of exogenous variation to estimate the tax competition parameter ($\lambda$) on first difference model. He estimates the following model:
 $$T_{i2000}-T_{i1999}=\lambda \sum_{j\neq i}w_{ij}(T_{j2000}-T_{j1999})+ \beta_0(X_{i2000}-X_{i1999})+ \beta_1 \sum_{j\neq i}w_{ij}(X_{j2000}-X_{j1999}) + v_i.$$
where $w_{ij}=1/n_{i}$ with $n_i$ the number of neighbor of the individual $i$.\\
The second column in Table 1 replicates the estimate using the instrument from \cite{lyytikainen2012tax}. He assumes that $\beta_1=0$ and use only one excluded instrument. Other estimations are carried out using spatial lag of the second-, third- and fourth-order and regularized estimators.\footnote{The instrument used in Table 3 of \cite{lyytikainen2012tax} is one of the instruments used with the spatial lag of other exogenous variables. I have augmented this model to account for an exogenous network model.}
\begin{table}[h!]
\centering
\caption{Estimates of the Tax Competition Parameter for Municipal Property Tax($n=411$)}
\label{Tax_comp}
\small{\begin{tabular}{lcccc}
 \toprule
Estimators/IVs &	\cite{lyytikainen2012tax}&	Spatial lags 2&	Spatial lags 2 and 3 &	Spatial lags 2, 3 and 4\\
\midrule
2SLS &	0.06 (0.07)&	0.26(0.28)& -0.02 (0.22)	&-0.03(0.17)\\
T-2SLS &0.01(0.004)	&	0.19(0.30)	&0.185(0.31)	&0.182(0.26)\\
L-2SLS & 0.01(0.0005)	&	0.20(0.22)&	0.186(0.33)&	0.182(0.31)\\
PC-2SLS &0.0115(0.42)	&	0.26(0.28)	&-0.027(0.22)	&-0.039(0.17)\\
Cond. number($\frac{\nu_1}{\nu_{min}}$) &	 2153.8&	2001	& 17800	&1.3983e+05\\
\bottomrule
\end{tabular}}
\tiny{Standard errors are in parenthesis. The change in general property taxation between 1999 and 2000 is the dependent variable. The independent variables are changes in neigboring municipalities' tax rates, the municipality's own imposed increase, non-zero own imposed increase and changes in municipal attributes, such as grants from the central government, disposable income per
capita, the unemployment rate and age structure (see Table 3 of \cite{lyytikainen2012tax}) for more details). The last line of the table shows the condition numbers of $QQ'$ matrices for different instrument sets. The values are relatively large, suggesting a near-perfect collinearity problem in small samples.}
\end{table}

The results in Table 1 suggest that the use of many instruments by adding more spatial lags biases the results. The use of regularization seems to reduce the bias of the estimation. The simulation results indicate that  T-2SLS and L-2SLS are the best methods in terms of bias correction. The point estimates obtained by both estimation methods are very similar, which suggests a bias correction relative to the 2SLS. As the number of instruments increases, the standard errors decrease for the 2SLS as well as for the regularized 2SLS. However, the standard errors are still very large, which means that the tax competition effect is not statistically significantly different from zero.\footnote{Inference using the standard errors of regularized estimators does not account for regularization and should be interpreted with caution, given the relatively small sample ($n=411$ municipalities).}

This empirical example shows how the regularized estimator can be used to improve the estimation of network models. The size of the tax competition parameter appears to be larger than is suggested by \cite{lyytikainen2012tax}. The estimators are not statistically different from zero. However, the regularized estimators (T-2SLS and L-2SLS) appear to be more stable as the number of instruments increases, which suggests that the weak identification problem may have been solved.

\section{Conclusion}
This paper uses regularization methods to estimate network models. It proposes easy-to-check  identification conditions based on the network adjacency matrix number of distinct eigenvalues.
Regularization is proposed as a solution to the weak identification problem in network models. Identification of the network effect can be achieved by using individuals' \cite{bonacich1987power} centrality as  instrumental variables. However, the number of instruments increases with the number of groups, leading to the many instruments problem.
Identification can also be achieved using the friend-of-a-friend's exogenous characteristics. However, if the network is very dense or group size is very large, the identification is weakened.

The proposed regularized 2SLS  estimators based on three regularization methods help address the  weak identification and many moments problems.
These estimators are consistent and asymptotically normal. The regularized 2SLS estimators achieve the asymptotic efficiency bound.
I derive an optimal data-driven selection method for the regularization parameter. An application to the estimation of tax competition in Finnish municipalities shows the empirical relevance of my methods.

A Monte Carlo experiment shows that the  regularized estimator performs well. The regularized  2SLS  procedures substantially
reduce the  bias from the 2SLS estimators, specifically in a large sample.  Moreover, the regularized estimator becomes more precise and less biased with increases in the network density and in the number of groups. These results show that regularization is a valuable solution to the potential weak identification problem existing in the estimation of network models.

\newpage
\appendix
\section{Appendix: Summary of notation}
To simplify notation, I use the following:

$P=P^{\alpha}$ , $q_j=q(\nu^2_j, \alpha)$ \\
$tr(A)$ is the trace of matrix $A$\\
$e_j$ is the $j^{th}$ unit (column) vector $j=1,...,n$\\
$e_f=\frac{1}{n}f'(I-P)f$ \\
$e_{2f}=\frac{1}{n}f'(I-P)^2f$,\\
$\Delta_f=tr(e_f)$ and $\Delta_{2f}=tr(e_{2f})$\\

\section{Appendix: Lemmas}
\textbf{Lemma 0: (Lemma 4 and Lemma 5 of Carrasco (2012))}\\
(i) $tr(P) = \sum_{j}q_j =O(1/\alpha)$ and $tr(P^2) = \sum_{j}q_j^2 =o((\sum_{j}q_j)^2)$, Lemma 4 (i) of Carrasco (2012),\\
(ii) $\Delta_{2f}=\left \{
  \begin{array}{r}
  O_p(\alpha^{\omega})\hspace{0.1cm} for \hspace{0.1cm} LF\hspace{0.1cm} and\hspace{0.1cm}SC\\
  O_p(\alpha^{min(\omega,2)})\hspace{0.1cm} for \hspace{0.1cm} T
  \end{array}%
  \right. $ and $f'(I-P)\varepsilon/\sqrt{n}=O_p(\sqrt{\Delta_{2f}})$, Lemma 5 (i) and (ii) of Carrasco (2012), \\
(iii) $u'P\varepsilon=O_p(1/\alpha)$, Lemma 5 (iii) of Carrasco (2012),\\
(iv) $E[u'P\varepsilon\varepsilon'Pu|X]=(\sum_jq_j)^2\sigma_{u\varepsilon}\sigma_{u\varepsilon}'+(\sum_jq_j^2)(\sigma_{u\varepsilon}\sigma_{u\varepsilon}'+\sigma_{\varepsilon}^2\Sigma_u)$, Lemma 5 (iv) of Carrasco (2012),\\
(v) $E[f'(I-P)\varepsilon\varepsilon'Pu/n|X]=O_p(\Delta_{2f}/\sqrt{\alpha n})$, Lemma 5 (viii) of Carrasco (2012).\\

\noindent \textbf{Lemma 1:}\\
(i) $tr(P) = \sum_{j}q_j =O(1/\alpha)$ and $tr(P^2) = \sum_{j}q_j^2 =o((\sum_{j}q_j)^2)$.\\
(ii) Suppose that $\{A\}$ is a sequence of
$n \times n $ UB matrices. For $B = PA$, $tr(B) = o((\sum_{j}q_j)^2)$, $tr(B^2) =  o((\sum_{j}q_j)^2)$, and
$\sum_i B_{ii}^2=o((\sum_{j}q_j)^2)$, where $B_{ii}$ are diagonal elements of $B$.

\noindent \textbf{Proof of Lemma 1:}\\
(i) Proof is in \cite{Carrasco12} Lemma 4 (i).\\
(ii) By eigenvalue decomposition, $AA'=\Pi \Delta \Pi'$, where $\Pi$ is an orthonormal
matrix and $ \Delta$ is the eigenvalue matrix. It follows that $PAA'P \leq \nu_{max} P^2$ with $\nu_{max}$ being the largest eigenvalue.
It follows that $tr(PAA'P) \leq \nu_{max} tr(P^2)=o_p((\sum_{j}q_j)^2)$. By the Cauchy-Schwarz inequality, $tr(B)\leq [ tr(P^2)]^{1/2}[tr(PAA'P)]^{1/2}= o_p((\sum_{j}q_j)^2)$.
 Also by the Cauchy-Schwarz inequality, $tr(B)\leq tr(BB')=tr(PAA'P)=o((\sum_{j}q_j)^2)$.\\

\noindent \textbf{Lemma 2:} Let $C$ and  $D$ be two UB $n \times n$  matrix sequences. \\
  (i) $C' P D=O_p(n/\alpha)$ \\
  (ii) $\varepsilon' C' P D\varepsilon=O_p(1/\alpha^2)$ and $C' P D\varepsilon=O_p(\sqrt{n}/\alpha)$

\noindent \textbf{Proof of Lemma 2:}\\
 (i) By the Cauchy-Schwarz inequality, $|e_i'C' P^{\alpha }D e_j| \leq \sqrt{e_i'C'C e_i} \sqrt{e_j'D' P^2D e_j} =O(n/\alpha)$, which implies that $C' PD=O(n/\alpha)$. \\
 (ii) $E|\varepsilon' C' P D\varepsilon|\leq \sqrt{E(\varepsilon'C'P^2 C \varepsilon)} \sqrt{E(\varepsilon'D' P^2D \varepsilon)}=\sigma^2 \sqrt{tr(C'P^2 C )} \sqrt{tr(D'P^2 D )}=O(\frac{1}{\alpha^2})$.\\
 By the Markov inequality, $\varepsilon' C' P D\varepsilon=O_p(\frac{1}{\alpha^2})$.\\
 By the Cauchy-Schwarz inequality, $|e_j' C' P D\varepsilon| \leq \sqrt{e_j'C'C e_j} \sqrt{\varepsilon'D' P^2D \varepsilon}=O_p(\sqrt{n}/\alpha)$, thus $C' P D\varepsilon=O_p(\sqrt{n}/\alpha)$.\\

\noindent \textbf{Lemma 3:}
 Suppose $\tilde{\rho}$ is a consistent estimator of $\rho_0$ and $\tilde{R} = R(\tilde{\rho})$.\\
 Then, $\frac{1}{n}Z^{\prime }\tilde{R}^{\prime } P\tilde{ R} Z=\frac{1}{n}Z^{\prime }R' P R Z+ O_p[(\tilde{\rho}-\rho_0)/\alpha]$ and \\$\frac{1}{n}Z^{\prime }\tilde{R}^{\prime } P\tilde{ R} R^{-1}\varepsilon =\frac{1}{n}Z^{\prime }R P\varepsilon + O_p[(\tilde{\rho}-\rho_0)/(\alpha\sqrt{n}(1+\alpha\sqrt{n}))]$.\\

\noindent \textbf{Proof of Lemma 3:}\\
 $\tilde{R}=R-(\tilde{\rho}-\rho_0)M$. Thus,
 \begin{eqnarray*}
Z^{\prime }\tilde{R}^{\prime } P\tilde{ R} Z/n&=&Z^{\prime }R' P R Z/n \\
 &-&(\tilde{\rho}-\rho_0)Z^{\prime }M' P R Z/n-(\tilde{\rho}-\rho_0)Z^{\prime }R' P M Z/n \\
 &+&(\tilde{\rho}-\rho_0)^2Z^{\prime }M' P M Z/n
\end{eqnarray*}%

  Let  us show that $Z^{\prime }R' P M Z/n=O_p(1/\alpha)$ and $Z^{\prime }M' P M Z/n=O_p(1/\alpha)$.\\
  Note that $Z=[WS^{-1}(X\beta_0+\iota \gamma_0), X]+WS^{-1}R^{-1}\varepsilon e_1'$.

  Under Assumption 3,
 $Z^{\prime }R' P M Z/n= O(1/\alpha)+O_p(1/\sqrt{n}\alpha)+ O_p(1/n\alpha^2)=O_p(1/\alpha)$
 and $Z^{\prime }M' P M Z/n= O_p(1/\alpha)$ by Lemma 2 (i).

\begin{eqnarray*}
Z^{\prime }\tilde{R}^{\prime } P\tilde{ R} \varepsilon/n&=&Z^{\prime }R P \varepsilon/n \\
 &-&(\tilde{\rho}-\rho_0)Z^{\prime }M' P \varepsilon/n-(\tilde{\rho}-\rho_0)Z^{\prime }R' P MR^{-1} \varepsilon/n \\
 &+&(\tilde{\rho}-\rho_0)^2Z^{\prime }M' P MR^{-1} \varepsilon/n
\end{eqnarray*}%

 Using the same argument as in the previous case under Assumption 3,

  $Z^{\prime }R' P M R^{-1}\varepsilon/n= O_p(1/\sqrt{n}\alpha+1/n\alpha^2)=O_p[1/\alpha\sqrt{n}(1+1/\alpha\sqrt{n})]$ ,
  $Z^{\prime }M' P \varepsilon/n=O_p[1/\alpha\sqrt{n}(1+1/\alpha\sqrt{n})]$ and $Z^{\prime }M' P \varepsilon/n=O_p[1/\alpha\sqrt{n}(1+1/\alpha\sqrt{n})]$
  by Lemma 2 (ii).\\

\noindent  \textbf{Lemma 4:} If Assumptions 1-4 are satisfied and $\alpha \rightarrow 0$, then\\
  (i)$Z^{\prime }R P R Z/n= H+ o_p(1)$ if $\alpha\sqrt{n} \rightarrow \infty$, and\\
  (ii)$Z^{\prime }R P \varepsilon/\sqrt{n}=f'\varepsilon/\sqrt{n} + o_p(1)$ if $\alpha^2\sqrt{n} \rightarrow \infty$.

\noindent  \textbf{Proof of Lemma 4:}\\
  Let $v=JRWS^{-1}R^{-1}\varepsilon$ and $JRZ=f+ve_1'$\\
  (i) $\frac{1}{n}Z^{\prime }R P R Z=\frac{1}{n} f'f-\frac{1}{n}f'(I-P)f+\frac{1}{n}e_1v'Pve_1'+ \frac{1}{n}f'Pve_1'+\frac{1}{n}e_1v'Pf$

  Let $e_f=\frac{1}{n}f'(I-P)f$, $e_{2f}=\frac{1}{n}f'(I-P)^2f$, $\Delta_f=tr(e_f)$ and $\Delta_{2f}=tr(e_{2f})$.
  By the Cauchy-Schwarz inequality,
  $\frac{1}{n}|e_i' f'(I-P)fe_j| \leq \frac{1}{n}\sqrt{e_i'f'f e_i} \sqrt{e_j'f'(I-P)^2f e_j}=O(\sqrt{\Delta_{2f}}).$

  From \cite{Carrasco12} Lemma 5 (i), $\Delta_{2f}=\left \{
  \begin{array}{r}
  O_p(\alpha^{\omega})\hspace{0.1cm} for \hspace{0.1cm} LF\hspace{0.1cm} and\hspace{0.1cm}SC\\
  O_p(\alpha^{min(\omega,2)})\hspace{0.1cm} for \hspace{0.1cm} T
  \end{array}%
  \right. $.
  Thus, $\Delta_{2f}=o_p(1)$.\\
  By Lemma 2 (ii), $\frac{1}{n}e_1v'Pve_1'+ \frac{1}{n}f'Pve_1'+\frac{1}{n}e_1v'Pf=O_p(\frac{1}{n\alpha^2}+\frac{1}{\alpha  \sqrt{n}})=o_p(1)$.\\
  (ii) $Z^{\prime }R P \varepsilon/\sqrt{n}=f'\varepsilon/\sqrt{n}-f'(I-P)\varepsilon/\sqrt{n}+ e_1v'P\varepsilon/\sqrt{n} $

  By Lemma 5 (ii) of \cite{Carrasco12}, $f'(I-P)\varepsilon/\sqrt{n}=O_p(\sqrt{\Delta_{2f}})$ and by Lemma 2 (ii), $e_1v'P\varepsilon/\sqrt{n}=O_p(1/\alpha^2\sqrt{n})$.

\section{Appendix: Proofs of propositions}
\textbf{Proof of Proposition 1:}\\
The Cayley-Hamilton theorem in linear algebra state that each square matrix is solution to it characteristic polynomial. The adjacency matrix of the network in our case is given by $W$, which is an $n \times n$ matrix.  If it has two distinct eigenvalues, therefore, the characteristic polynomial, $p(\tau)=det(\tau I_n-W)$,  is a degree two polynomial.  Thus, there exist $a_0, a_1$ and, $a_2$ with $a_2\neq 0$ such that $a_0I_n+ a_1W+a_2W^2=0$. $I_n$, $W$ and $W^2$ are linearly dependant and from Proposition 1 of \cite{bramoulle2009identification} the network effects are not identified.

\textbf{Proof of Proposition 2:}\\
Under Assumption 2 (i.e. that $Sup \|\lambda W \|<1$), $f$ can be approximated by a linear combination of $(JWX, JW^2X,..., JW^{\varrho_w-1}X)$  and $JX$. Indeed, using Caley-Hamilton theorem and the fact that the characteristic polynomial has $\varrho_w$ distinct eigenvalues, For any natural number $q>\varrho_w$,  $W^q$ can be written as a linear combination of $I_n, W,...,W^{\varrho_w-1}$. Thus, $WS^{-1}$ can be written a linear combination of $I_n, W,...,W^{\varrho_w-1}$. Therefore,  $f$ can be approximated by a linear combination of $(JWX, JW^2X,..., JW^{\varrho_w-1}X)$  and $JX$.

Let  assume that $[WX, W^2X, . . ., W^{\varrho_w-1}X, X ]$  is full rank column.\newline Let $Q=J[WX, W^2X, . . ., W^{\varrho_w-1}X, X]$ be the set of instrumental variables. The identification of the network effects is based on the moment conditions $E(Q'\varepsilon(\rho_0, \delta))=0$ (i.e. $Q'f(\delta_0-\delta)=0$). The parameters are point identified if the solution to this equation is unique.  A necessary and sufficient condition is that $Q$ and $f$ are  full rank column. $[WX, W^2X, . . ., W^{\varrho_w-1}X, X ]$  is full rank column if and only if  $Q$  is full rank column. Moreover, if $[WX, W^2X, . . ., W^{\varrho_w-1}X, X ]$  is full rank column the $f$ is of rank $1+k$.

Let assume that $[WX, W^2X, . . ., W^{\varrho_w-1}X, X]$ is not full rank column. Consider $$\mathfrak{B}=\{ b\in \mathbb{R}^{k\times\varrho_w},X b_0+WX b_2+. . .+ W^{\varrho_w-1}X b_{\varrho_w-1}=0\}$$ It can be observed that $f=[JWS^{-1}(X\beta_0),  JX]$ is equivalent to $f=J[\sum_{k=1}^{\varrho_w-1}\varsigma_k W^kX\beta_0,  X]$. Consider $$\mathfrak{A}=\{ a=(a_0, a_1)\in \mathbb{R}^{k}\times \mathbb{R}, Xa_0+ a_1\sum_{k=1}^{\varrho_w-1}\varsigma_k W^kX\beta_0  =0 \}$$
$f$ is not full rank if and only if $\mathfrak{A} \neq \{0\}$.

In other word, $f$ is not full rank column if and only  if there exist $b \in \mathfrak{B}$ such that $b_0=a_0$, $b_k=a_1\varsigma_k\beta_0$ with $\beta_0$, $\varsigma_k$ known constant for all $k=1,...,\varrho_w-1$ and $b\neq0$. The condition for $f$ not  being full rank column of very restrictive. However, if we assume that there exist such a sub set in $\mathfrak{A}$, then $f$ is not full rank.

 Note that in general, it is possible to have $JWS^{-1}(X\beta_0+\iota\gamma_0)$ linearly independent from $JX$ without $[WX, W^2X, . . ., W^{\varrho_w-1}X, X ]$  being full rank column. This happen if  $\beta_0$, $\lambda$ and $\gamma_0$ are not in the space parameter compatible with the null space of $[WX, W^2X, . . ., W^{\varrho_w-1}X, X ]$.

The condition $[WX, W^2X, . . ., W^{\varrho_w-1}X, X ]$ is full rank column is therefore a necessary but not, in general, a sufficient condition for identification. But if we restrict the true value of the parameter to be in the compatible set as in \cite{bramoulle2009identification} Result 1 (2) Page 54 the condition is necessary and sufficient.


\textbf{Proof of Proposition 3:}\\
The proof of proposition 3 is similar to that of proposition 2 with $[WX, W^2X, . . ., W^{\varrho_w-1}X, X]$ replaced by $Q_{0\varrho_w}=[WX, W^2X, . . .W^{\varrho_w-1}X, W\iota, W^2\iota, . . .,W^{\varrho_w-1}\iota, X]$. The identification result in this case are conditional on a consistent preliminary estimation of $\rho$ as in \cite{LeeandLung2010Central}.

  \textbf{Proof of Proposition 4:}\\
  The regularized 2SLS estimator satisfies
  $ \hat{\delta}_{R2sls}-\delta_0=(Z^{\prime }\tilde{R}^{\prime } P\tilde{ R} Z)^{-1}Z^{\prime }\tilde{R}^{\prime } P \tilde{ R} R^{-1}\varepsilon.$\\
  $Z^{\prime }\tilde{R}^{\prime } P\tilde{ R} Z/n=O_p(1)+O_n(1/\alpha \sqrt{n})$ by Lemmas 3 and 4. \\
  $\tilde{R}^{\prime } P \tilde{ R} R^{-1}\varepsilon/n =O_p(1/\sqrt{n})+O_p[(1/(n\alpha (1+\alpha\sqrt{n}))$ by Lemmas 3 and 4.

  Then, $ \hat{\delta}_{R2sls}-\delta_0=o_p(1)$ as  $\alpha\sqrt{n} \rightarrow \infty$ and $\alpha \rightarrow 0$.
  This proves the consistency of the regularized 2SLS for SAR with many instruments:
   $$ \sqrt{n}(\hat{\delta}_{R2sls}-\delta_0)=(Z^{\prime }\tilde{R}^{\prime } P\tilde{ R} Z/n)^{-1}[Z^{\prime }\tilde{R}^{\prime } P \tilde{ R} R^{-1}\varepsilon/\sqrt{n}].$$\\
   Using Lemmas 3 and 4, as well as the  Slutzky theorem:
   $$\sqrt{n}(\hat{\delta}_{R2sls}-\delta_0)\overset{d}{\rightarrow }\mathcal{N} \left( 0,\sigma _{\varepsilon }^{2}H^{-1}\right) $$
   if $\alpha^2\sqrt{n} \rightarrow \infty$ and $\alpha \rightarrow 0$.
   \newline

 \noindent  \textbf{Proof of Proposition 5 }\\

 \noindent  Let us consider the MSE of the estimated parameters:
   $$n(\hat{\delta}_{R2sls}-\delta_0)(\hat{\delta}_{R2sls}-\delta_0)=\hat{H}^{-1}\hat{h}\hat{h}'\hat{H}^{-1}$$
   with $\hat{H}=\frac{Z'\tilde{R}' P \tilde{R} Z}{n}$ and $\hat{h}=\frac{Z'\tilde{R}'P\tilde{R}Y}{\sqrt{n}}$. Our objective is to approximate the MSE.
   To achieve this, I use a Nagar-type approximation in order to concentrate on the largest part of the MSE.
  By Lemma 3,\begin{eqnarray*}
               \hat{H} &=& Z^{\prime }R P R Z/n \\
 &-&(\tilde{\rho}-\rho_0)Z^{\prime }M' P R Z/n-(\tilde{\rho}-\rho_0)Z^{\prime }R' P M Z/n \\
 &+&(\tilde{\rho}-\rho_0)^2Z^{\prime }M' P M Z/n.
  \end{eqnarray*}
  And $ \hat{H}= Z^{\prime }R P R Z/n+ O_p((\tilde{\rho}-\rho_0)/\alpha)$. By Lemma 4, we have that

$$\hat{H}=\frac{1}{n} f'f-\frac{1}{n}f'(I-P)f+\frac{1}{n}e_1v'Pve_1'+ \frac{1}{n}f'Pve_1'+\frac{1}{n}e_1v'Pf + O_p((\tilde{\rho}-\rho_0)/\alpha).$$
Let us define  $T^H=T^H_1+ T^H_2+ T^H_3$, with $T^H_1= -\frac{1}{n}f'(I-P)f$, $T^H_2= \frac{1}{n}e_1v'Pve_1'$ and $T_3^H= \frac{1}{n}f'Pve_1'+\frac{1}{n}e_1v'Pf + O_p((\tilde{\rho}-\rho_0)/\alpha) $,
such that \begin{eqnarray*}
            \hat{H} &=& \frac{1}{n} f'f+T_1^H+ T_2^H+ T_3^H \\
             &=& H+ T_1^H+ T_2^H+ T_3^H+o_p(1) \\
             &=&  H+T^H+o_p(1).
          \end{eqnarray*}

Following similar arguments, we have $$\hat{h}=f'\varepsilon/\sqrt{n}-f'(I-P)\varepsilon/\sqrt{n}+ e_1v'P\varepsilon/\sqrt{n} + O_p[(\tilde{\rho}-\rho_0)/(\alpha(1+\alpha\sqrt{n}))].$$
Let us also define $T^h=T^h_1+T^h_2$  with\\
$T^h_1=-f'(I-P)\varepsilon/\sqrt{n}$ and $T^h_2= e_1v'P\varepsilon/\sqrt{n} + O_p[(\tilde{\rho}-\rho_0)/(\alpha(1+\alpha\sqrt{n}))]$.

We therefore have \begin{eqnarray*}
                    \hat{h} &=& f'\varepsilon/\sqrt{n}+ T^h_1+ T^h_2 \\
                     &=& h+ T^h_1+ T^h_2+o_p(1) \\
                    &=& h+T^h+o_p(1).
                  \end{eqnarray*}

Using a Nagar-type expansion on $\hat{H}^{-1}$,
 $$n(\hat{\delta}_{R2sls}-\delta_0)(\hat{\delta}_{R2sls}-\delta_0)=H^{-1}{[I-T^HH^{-1}][hh'+hT^h+T^hh'+T^hT^{h'}][I-H^{-1}T^H]}H^{-1}+ o_p(1).$$

   Let us define $A(\alpha)=[I-T^HH^{-1}]\Im(\alpha) [I-H^{-1}T^H]$ with $\Im(\alpha)=[hh'+hT^h+T^hh'+T^hT^{h'}]$.

   Therefore,
   $A(\alpha)= \Im(\alpha)+T^HH^{-1}\Im(\alpha)H^{-1}T^H- T^HH^{-1}\Im(\alpha)-\Im(\alpha)H^{-1}T^H$.

   \begin{eqnarray*}
     E[\Im(\alpha)|X] &=& \sigma^2 [H-2e_f+ \frac{1}{n} f'Pve_1'+\frac{1}{n} e_1 v'P f+e_{2f}] \\
      &-& E[\frac{1}{n} f'(I-P)\varepsilon \varepsilon' P ve_1'+\frac{1}{n} e_1 v'P\varepsilon \varepsilon'(I-P)f|X] \\
      &+& E[\frac{1}{n} e_1 v'P\varepsilon \varepsilon'P ve_1'|X].
   \end{eqnarray*}

   $E(T^HH^{-1}\Im(\alpha)|X)=-\sigma^2 e_f+ o_p(1)$ and $E(\Im(\alpha)H^{-1}T^H|X)=-\sigma^2 e_f +o_p(1)$.\\
   \begin{eqnarray*}
      E(T^HH^{-1}\Im(\alpha)H^{-1}T^H|X)&=&  \sigma^2 HO_p([\frac{1}{n\alpha^2}+\frac{1}{\alpha  \sqrt{n}}+\Delta_{f}]^2)\\
      &=&  O_p([\frac{1}{n\alpha^2}+\frac{1}{\alpha  \sqrt{n}}+\Delta_{f}]^2).
   \end{eqnarray*}

   We have
   \begin{eqnarray*}
   E( A(\alpha)|X) &=& \sigma^2H + \sigma^2e_{2f}+E[\frac{1}{n} e_1 v'P\varepsilon \varepsilon'P ve_1'|X] \\
    &-& E[\frac{1}{n} f'(I-P)\varepsilon \varepsilon' P ve_1'+\frac{1}{n} e_1 v'P\varepsilon \varepsilon'(I-P)f|X] \\
   &+& \frac{1}{n} f'Pve_1'+\frac{1}{n} e_1 v'P f+ O_p([\frac{1}{n\alpha^2}+\frac{1}{\alpha \sqrt{n}}+\Delta_{f}]^2).
   \end{eqnarray*}

   From Lemma 5 (viii) of Carrasco (2012), we have $$E[\frac{1}{n} f'(I-P)\varepsilon \varepsilon' P ve_1'+\frac{1}{n} e_1 v'P\varepsilon \varepsilon'(I-P)f|X]=O_p(\sqrt{\Delta_{2f}}/\sqrt{\alpha n})$$ and $\frac{1}{n} e_1 v'(P-P^2)f=O_p(\sqrt{\Delta_{2f}}/\sqrt{\alpha n})$.

  From Lemma 5 (iii) of Carrasco (2012), $\frac{1}{n} f'Pve_1'+\frac{1}{n} e_1 v'P f= O_p(\frac{1}{n\alpha})$.

  And, from Lemma 5 (iv) of Carrasco (2012), $$E[\frac{1}{n} e_1 v'P\varepsilon \varepsilon'P ve_1'/n|X]=\frac{1}{n} (\sum_jq_j)^2 \sigma^4e_1\iota'D'D\iota e_1' + o_p((\sum_jq_j)^2/n)$$  with $D=JRWS^{-1}R^{-1}$.

  We can conclude that $$n(\hat{\delta}_{R2sls}-\delta_0)(\hat{\delta}_{R2sls}-\delta_0)= Q(\alpha) + \hat{R}(\alpha)$$
  with $E[Q(\alpha)| X]=H^{-1}\sigma^2+ H^{-1}\left[ \sigma^2e_{2f}+\frac{1}{n}(\sum_jq_j)^2 \sigma^4e_1\iota'D'D\iota e_1' \right]H^{-1}$ and $$r(\alpha)=E(\hat{R}(\alpha)|X)= o_p((\sum_jq_j)^2/n)+ O_p([\frac{1}{n\alpha^2}+\frac{1}{\alpha  \sqrt{n}}+\Delta_{f}]^2+\frac{1}{n\alpha}+\frac{\Delta_{2f}}{\sqrt{\alpha n}}).$$

$S(\alpha)= H^{-1}\left[ \sigma^2e_{2f}+\frac{1}{n}(\sum_jq_j)^2 \sigma^4e_1\iota'D'D\iota e_1' \right]H^{-1}$.

  Note that $r(\alpha)/tr(S(\alpha))=o_p(1)$; my argument is similar to that used in Carrasco (2012). This means that $S(\alpha)$ is the dominant part of the MSE of the estimation of the model using regularized 2SLS.

\newpage
\bibliography{bibALL}
\bibliographystyle{econometrica}

\newpage

\section{Appendix: Monte Carlo Simulation Results}
Mean, standard deviation (SD) and root mean
square errors (RMSE) of the empirical distributions of the estimates are reported. Each data-generating process uses 500 replications.
\begin{table}[htbp]
\centering
  \caption{Simulation results with maximum of three connections (1/2)}
   \tiny{ \centering{\begin{tabular}{lccccc}
    \toprule
    $m=10$  &       & $g=30$  &       &       &  \\
    \midrule
    &       & $\lambda_0=0.1$ & $\beta_{10}=0.2$ & $\beta_{20}=0.2$ & $\rho_0=0.1$ \\
    \midrule
      2SLS (finite iv)  &       & 0.098 (0.207) [0.207] & 0.200 (0.071) [0.071] & 0.208 (0.071) [0.071] & 0.128 (0.231) [0.233] \\
    2SLS (large iv)  &       & 0.015 (0.100) [0.131] & 0.190 (0.068) [0.068] & 0.220 (0.060) [0.063] & - \\
    Bias-corrected 2SLS &       & 0.106 (0.131) [0.131] & 0.198 (0.069) [0.069] & 0.205 (0.063) [0.064] & - \\
    T-2SLS &       & 0.040 (0.110) [0.125] & 0.187 (0.079) [0.080] & 0.216 (0.065) [0.066] & - \\
    LF-2SLS &       & 0.052 (0.121) [0.130] & 0.188 (0.083) [0.084] & 0.215 (0.067) [0.068] & - \\
    PC-2SLS &       & 0.052 (0.121) [0.130] & 0.188 (0.083) [0.084] & 0.215 (0.067) [0.068] &  -\\
     \midrule
          &       & $g=60$  &       &       &  \\
     \midrule
      2SLS (finite iv)  &       & 0.104 (0.136) [0.136] & 0.203 (0.047) [0.047] & 0.204 (0.049) [0.049] & 0.116 (0.177) [0.178] \\
    2SLS (large iv)  &       & 0.032 (0.081) [0.105] & 0.196 (0.046) [0.046] & 0.217 (0.043) [0.046] & - \\
    Bias-corrected 2SLS &       & 0.108 (0.099) [0.099] & 0.202 (0.047) [0.047] & 0.204 (0.045) [0.045] & - \\
    T-2SLS &       & 0.055 (0.088) [0.099] & 0.193 (0.051) [0.052] & 0.213 (0.046) [0.048] & - \\
    LF-2SLS &       & 0.064 (0.095) [0.101] & 0.193 (0.053) [0.054] & 0.212 (0.048) [0.049] & - \\
    PC-2SLS &       & 0.064 (0.095) [0.101] & 0.193 (0.053) [0.054] & 0.212 (0.048) [0.049] &  -\\
    \bottomrule
    \end{tabular}%
  \label{tab:addlabel}}}\\%
  \scriptsize{Mean (SD) [RMSE]}
\end{table}%

\begin{table}[htbp]
\centering
  \caption{Simulation results with maximum of three connections (2/2)}
    \tiny{\centering{ \begin{tabular}{lccccc}
    \toprule
    $m=15$  &       & $g=30$  &       &       &  \\
    \midrule
    &       & $\lambda_0=0.1$ & $\beta_{10}=0.2$ & $\beta_{20}=0.2$ & $\rho_0=0.1$ \\
    \midrule
     2SLS (finite iv)  &       & 0.098 (0.155) [0.155] & 0.203 (0.052) [0.052] & 0.202 (0.055) [0.055] & 0.115 (0.204) [0.205] \\
    2SLS (large iv)  &       & 0.069 (0.094) [0.099] & 0.200 (0.052) [0.052] & 0.209 (0.044) [0.045] & - \\
    Bias-corrected 2SLS &       & 0.101 (0.105) [0.105] & 0.202 (0.052) [0.052] & 0.202 (0.047) [0.047] & - \\
    T-2SLS &       & 0.086 (0.106) [0.107] & 0.199 (0.059) [0.059] & 0.207 (0.049) [0.050] & - \\
    LF-2SLS &       & 0.089 (0.114) [0.115] & 0.200 (0.062) [0.062] & 0.207 (0.053) [0.054] & - \\
    PC-2SLS &       & 0.089 (0.114) [0.115] & 0.200 (0.062) [0.062] & 0.207 (0.053) [0.054] &  -\\
     \midrule
          &       & $g=60$  &       &       &  \\
 \midrule
    2SLS (finite iv)  &       & 0.093 (0.103) [0.103] & 0.198 (0.037) [0.037] & 0.200 (0.039) [0.039] & 0.109 (0.143) [0.143] \\
    2SLS (large iv)  &       & 0.066 (0.061) [0.070] & 0.197 (0.038) [0.038] & 0.206 (0.034) [0.035] & - \\
    Bias-corrected 2SLS &       & 0.096 (0.072) [0.072] & 0.198 (0.038) [0.038] & 0.199 (0.036) [0.036] & - \\
    T-2SLS &       & 0.079 (0.069) [0.072] & 0.196 (0.043) [0.043] & 0.204 (0.037) [0.037] & - \\
    LF-2SLS &       & 0.082 (0.074) [0.076] & 0.196 (0.046) [0.046] & 0.203 (0.040) [0.040] &  -\\
    PC-2SLS &       & 0.082 (0.074) [0.076] & 0.196 (0.046) [0.046] & 0.203 (0.040) [0.040] & - \\
    \bottomrule
    \end{tabular}%
  \label{tab:addlabel}}}\\%
  \scriptsize{Mean (SD) [RMSE]}
\end{table}%

\begin{table}[htbp]
\centering
  \caption{Simulation results with maximum of six connections (1/2)}
    \tiny{ \centering{\begin{tabular}{lccccc}
    \toprule
    $m=10$  &       & $g=30$  &       &       &  \\
    \midrule
   &       & $\lambda_0=0.1$ & $\beta_{10}=0.2$ & $\beta_{20}=0.2$ & $\rho_0=0.1$ \\
    \midrule
      2SLS (finite iv)  &       & 0.102 (0.118) [0.118] & 0.196 (0.074) [0.074] & 0.206 (0.049) [0.049] & 0.103 (0.187) [0.188] \\
    2SLS (large iv)  &       & 0.052 (0.056) [0.074] & 0.183 (0.069) [0.071] & 0.206 (0.047) [0.047] & - \\
    Bias-corrected 2SLS &       & 0.109 (0.078) [0.078] & 0.196 (0.072) [0.072] & 0.202 (0.048) [0.048] & - \\
    T-2SLS &       & 0.065 (0.063) [0.073] & 0.172 (0.079) [0.083] & 0.203 (0.051) [0.051] &  -\\
    LF-2SLS &       & 0.071 (0.069) [0.075] & 0.167 (0.085) [0.091] & 0.203 (0.053) [0.053] & - \\
    PC-2SLS &       & 0.071 (0.069) [0.075] & 0.167 (0.085) [0.091] & 0.203 (0.053) [0.053] &  -\\
     \midrule
          &       & $g=60$  &       &       &  \\
  \midrule
      2SLS (finite iv)  &       & 0.099 (0.090) [0.090] & 0.204 (0.051) [0.051] & 0.207 (0.034) [0.035] & 0.118 (0.158) [0.159] \\
    2SLS (large iv)  &       & 0.053 (0.038) [0.060] & 0.193 (0.047) [0.048] & 0.209 (0.032) [0.034] &  -\\
    Bias-corrected 2SLS &       & 0.099 (0.064) [0.064] & 0.203 (0.049) [0.049] & 0.205 (0.034) [0.034] &  -\\
    T-2SLS &       & 0.066 (0.044) [0.056] & 0.184 (0.054) [0.056] & 0.205 (0.035) [0.035] & - \\
    LF-2SLS &       & 0.072 (0.049) [0.057] & 0.180 (0.058) [0.061] & 0.204 (0.036) [0.036] & - \\
    PC-2SLS &       & 0.072 (0.049) [0.057] & 0.180 (0.058) [0.061] & 0.204 (0.036) [0.036] &  -\\
    \bottomrule
    \end{tabular}%
  \label{tab:addlabel}}}\\%
  \scriptsize{Mean (SD) [RMSE]}
\end{table}%

\begin{table}[htbp]
\centering
  \caption{Simulation results with maximum of six connections (2/2)}
    \tiny{ \centering{ \begin{tabular}{lccccc}
    \toprule
    $m=15$  &       & $g=30$  &       &       &  \\
    \midrule
    &       & $\lambda_0=0.1$ & $\beta_{10}=0.2$ & $\beta_{20}=0.2$ & $\rho_0=0.1$ \\
    \midrule
      2SLS (finite iv)  &       & 0.104 (0.068) [0.069] & 0.205 (0.053) [0.053] & 0.203 (0.038) [0.038] & 0.116 (0.144) [0.145] \\
    2SLS (large iv)  &       & 0.081 (0.043) [0.047] & 0.199 (0.052) [0.052] & 0.207 (0.035) [0.036] & - \\
    Bias-corrected 2SLS &       & 0.100 (0.065) [0.065] & 0.203 (0.053) [0.053] & 0.204 (0.037) [0.037] &  -\\
    T-2SLS &       & 0.092 (0.048) [0.049] & 0.198 (0.060) [0.060] & 0.207 (0.039) [0.040] &  -\\
    LF-2SLS &       & 0.095 (0.053) [0.053] & 0.198 (0.062) [0.062] & 0.207 (0.041) [0.041] & - \\
    PC-2SLS &       & 0.095 (0.053) [0.053] & 0.198 (0.062) [0.062] & 0.207 (0.041) [0.041] &  -\\
     \midrule
          &       & $g=60$  &       &       &  \\
 \midrule
    2SLS (finite iv)  &       & 0.103 (0.050) [0.050] & 0.200 (0.039) [0.039] & 0.200 (0.026) [0.026] & 0.108 (0.100) [0.100] \\
    2SLS (large iv)  &       & 0.086 (0.030) [0.033] & 0.196 (0.039) [0.039] & 0.202 (0.024) [0.025] & - \\
    Bias-corrected 2SLS &       & 0.105 (0.035) [0.035] & 0.200 (0.039) [0.039] & 0.199 (0.025) [0.025] & - \\
    T-2SLS &       & 0.094 (0.035) [0.035] & 0.193 (0.043) [0.043] & 0.201 (0.027) [0.027] &  -\\
    LF-2SLS &       & 0.097 (0.038) [0.039] & 0.193 (0.044) [0.044] & 0.201 (0.028) [0.028] & - \\
    PC-2SLS &       & 0.097 (0.038) [0.039] & 0.193 (0.044) [0.044] & 0.201 (0.028) [0.028] & - \\
    \bottomrule
    \end{tabular}%
  \label{tab:addlabel}}}\\%
  \scriptsize{Mean (SD) [RMSE]}
\end{table}%

\begin{table}[htbp]
\centering
  \caption{Simulation results with maximum of eight connections (1/2)}
   \tiny{\centering{ \begin{tabular}{lccccc}
    \toprule
    $m=10$  &       & $g=30$  &       &       &  \\
    \midrule
   &       & $\lambda_0=0.1$ & $\beta_{10}=0.2$ & $\beta_{20}=0.2$ & $\rho_0=0.1$ \\
    \midrule
    2SLS (finite iv)  &       & 0.092 (0.108) [0.108] & 0.191(0.073) [0.074] & 0.204 (0.047) [0.047] & 0.111 (0.211) [0.211] \\
    2SLS (large iv)  &       & 0.064 (0.043) [0.056] & 0.188 (0.069) [0.070] & 0.206 (0.045) [0.046] & - \\
    Bias-corrected 2SLS &       & 0.099 (0.062) [0.062] & 0.194 (0.071) [0.071] & 0.203 (0.048) [0.048] & - \\
    T-2SLS &       & 0.073 (0.049) [0.056] & 0.180 (0.083) [0.086] & 0.201 (0.049) [0.049] &  -\\
    LF-2SLS &       & 0.077 (0.054) [0.058] & 0.177 (0.093) [0.096] & 0.200 (0.051) [0.051] & - \\
    PC-2SLS &       & 0.077 (0.054) [0.058] & 0.177 (0.093) [0.096] & 0.200 (0.051) [0.051] &  -\\
     \midrule
          &       & g=60  &       &       &  \\
     \midrule
    2SLS (finite iv)  &       & 0.096 (0.065) [0.065] & 0.202 (0.048) [0.048] & 0.204 (0.033) [0.033] & 0.113 (0.162) [0.162] \\
    2SLS (large iv)  &       & 0.071 (0.028) [0.040] & 0.198 (0.047) [0.047] & 0.207 (0.032) [0.032] &  -\\
    Bias-corrected 2SLS &       & 0.102 (0.039) [0.039] & 0.202 (0.048) [0.048] & 0.202 (0.033) [0.033] &  -\\
    T-2SLS &       & 0.080 (0.032) [0.037] & 0.194 (0.056) [0.057] & 0.203 (0.034) [0.034] & - \\
    LF-2SLS &       & 0.084 (0.036) [0.039] & 0.192 (0.063) [0.063] & 0.201 (0.035) [0.035] & - \\
    PC-2SLS &       & 0.084 (0.036) [0.039] & 0.192 (0.063) [0.063] & 0.201 (0.035) [0.035] &  -\\
    \bottomrule
    \end{tabular}%
  \label{tab:addlabel}}}\\
\scriptsize{Mean (SD) [RMSE]}
\end{table}%

\begin{table}[htbp]
\centering
  \caption{Simulation results with maximum of eight connections (2/2)}
    \tiny{ \centering{\begin{tabular}{lccccc}
    \toprule
    $m=15$  &       & $g=30$  &       &       &  \\
    \midrule
    &       & $\lambda_0=0.1$ & $\beta_{10}=0.2$ & $\beta_{20}=0.2$ & $\rho_0=0.1$ \\
    \midrule
     2SLS (finite iv)  &       & 0.102 (0.052) [0.052] & 0.203 (0.053) [0.053] & 0.203 (0.033) [0.033] & 0.112 (0.136) [0.137] \\
    2SLS (large iv)  &       & 0.087 (0.028) [0.031] & 0.198 (0.051) [0.051] & 0.204 (0.032) [0.032] & - \\
    Bias-corrected 2SLS &       & 0.101 (0.036) [0.036] & 0.202 (0.052) [0.052] & 0.202 (0.032) [0.032] & - \\
    T-2SLS &       & 0.093 (0.032) [0.033] & 0.195 (0.058) [0.058] & 0.204 (0.034) [0.034] & - \\
    LF-2SLS &       & 0.095 (0.036) [0.036] & 0.193 (0.061) [0.061] & 0.204 (0.035) [0.035] & - \\
    PC-2SLS &       & 0.095 (0.036) [0.036] & 0.193 (0.061) [0.061] & 0.204 (0.035) [0.035] & - \\
     \midrule
          &       & $g=60$  &       &       &  \\
 \midrule
   2SLS (finite iv)  &       & 0.103 (0.038) [0.038] & 0.199 (0.039) [0.039] & 0.199 (0.022) [0.022] & 0.105 (0.097) [0.097] \\
    2SLS (large iv)  &       & 0.091 (0.020) [0.022] & 0.195 (0.038) [0.039] & 0.200 (0.021) [0.021] & - \\
    Bias-corrected 2SLS &       & 0.104 (0.026) [0.026] & 0.199 (0.039) [0.039] & 0.198 (0.021) [0.021] & - \\
    T-2SLS &       & 0.096 (0.023) [0.023] & 0.191 (0.043) [0.043] & 0.200 (0.022) [0.022] & - \\
    LF-2SLS &       & 0.098 (0.026) [0.026] & 0.190 (0.044) [0.045] & 0.200 (0.023) [0.023] & - \\
    PC-2SLS &       & 0.098 (0.026) [0.026] & 0.190 (0.044) [0.045] & 0.200 (0.023) [0.023] & - \\
    \bottomrule
    \end{tabular}%
  \label{tab:addlabel}}}\\
  \scriptsize{Mean (SD) [RMSE]}
\end{table}%

\end{document}